\documentclass[prb,aps,notitlepage,nofootinbib,superscriptaddress,twocolumn,longbibliography,10pt]{revtex4-2}
\usepackage[normalem]{ulem}
\usepackage{svg}
\usepackage{amsmath}
\usepackage{amsthm, amssymb}
\usepackage[colorlinks=true,citecolor=blue,linkcolor=blue]{hyperref}
\usepackage{graphicx}
\usepackage{dcolumn}
\usepackage{bm}
\usepackage{color}
\usepackage{tabularx}
\usepackage{comment}
\usepackage{mathtools}
\usepackage{tikz-cd}
\usepackage{adjustbox}
\usepackage{braket}
\usepackage{dsfont}
\usepackage{subcaption}
\usepackage{ragged2e}
\usepackage[linesnumbered,ruled,vlined]{algorithm2e}
\SetKwComment{Comment}{/* }{ */}
\SetKwInput{Input}{Input}
\SetKwInput{Return}{Return}
\SetKwFor{For}{for}{do}{endfor}
\DontPrintSemicolon
\usepackage{tikz}
\usetikzlibrary{quantikz2}

\newcommand{\Tr}{\text{Tr}}
\begin{document}

\begin{titlepage}

\title{Postselection-free experimental observation of the measurement-induced phase transition in circuits with universal gates}
\author{Xiaozhou Feng}

\affiliation{Department of Physics, The University of Texas at Austin, Austin, Texas 78712, USA}
\author{Jeremy C\^{o}t\'{e}}

\author{Stefanos Kourtis}
\affiliation{Institut quantique \& D\'{e}partement de physique, Universit\'{e} de Sherbrooke, Sherbrooke (Qu\'{e}bec) J1K 2R1, Canada}
\author{Brian Skinner}
\affiliation{Department of Physics, The Ohio State University, Columbus, Ohio 43202, USA}

\setcounter{equation}{0}
\setcounter{figure}{0}
\setcounter{table}{0}

\makeatletter
\renewcommand{\theequation}{S\arabic{equation}}
\renewcommand{\thefigure}{S\arabic{figure}}
\renewcommand{\thetable}{S\Roman{table}}
\renewcommand{\bibnumfmt}[1]{[S#1]}
\renewcommand{\citenumfont}[1]{S#1}

\date{\today}
\begin{abstract}
   Monitored many-body systems can exhibit a phase transition between entangling and disentangling dynamical phases by tuning the strength of measurements made on the system as it evolves. This phenomenon is called the measurement-induced phase transition (MIPT). Understanding the properties of the MIPT is a prominent challenge for both theory and experiment at the intersection of many-body physics and quantum information. 
   Realizing the MIPT experimentally is particularly challenging due to the postselection problem, which demands a number of experimental realizations that grows exponentially with the number of measurements made during the dynamics. Proposed approaches that circumvent the postselection problem typically rely on a classical decoding process that infers the final state based on the measurement record. But the complexity of this classical process generally also grows exponentially with the system size unless the dynamics is restricted to a fine-tuned set of unitary operators. In this work we overcome these difficulties. We construct a tree-shaped quantum circuit whose nodes are Haar-random unitary operators followed by weak measurements of tunable strength. For these circuits, we show that the MIPT can be detected without postselection using only a simple classical decoding process whose complexity grows linearly with the number of qubits. Our protocol exploits the recursive structure of tree circuits, which also enables a complete theoretical description of the MIPT, including an exact solution for its critical point and scaling behavior. We experimentally realize the MIPT on Quantinuum's H1-1 trapped-ion quantum computer and show that the experimental results are precisely described by theory. Our results close the gap between analytical theory and postselection-free experimental observation of the MIPT.
\end{abstract}
\maketitle
\vspace{2mm}
\end{titlepage}

\section{Introduction} \label{sec:introduction}

When a many-body quantum system undergoes unitary evolution that is subject to sporadic measurements, its dynamics falls into one of two broad classes, generally called the entangling and disentangling dynamical phases. Separating these two phases is a measurement-induced phase transition (MIPT)~\cite{Skinner_2019,Li_2018,Chan_2019}. The nature of this transition has been intensely studied in recent years (e.g., refs.~\cite{Li_Measurement-driven_2019,Choi_2020,Gullans_2020,Bao_Theory_2020,Jian_Measurement_2020,Li_Conformal_2021,Zabolo_Critical_2020,Gullans_Scalable_2020,Tang_Measurement-induced_2020,Turkeshi_Measurement-induced_2020,Fan_Self-organized_2021,Li_Statistical_2021,nahum_measurement_2021,li2021statistical,Lavasani_Measurement_2021,Ippoliti_Entanglement_2021,Lopez-Piqueres_Mean-field_2020,Noel_2022,Ippoliti_postselection_2021,Li_Cross_2022,li_entanglement_2023,feng_measurement_2023,Hoke_Measurement-induced_2023,Fava_Nonlinear_2023,jian2023measurementinduced,agrawal_entanglement_2022,learnability_Fergus,majidy_critical_2023,koh2023measurement,Ippoliti_Learnability_2024,Akhtar_Measurement_2024,garratt2023probing,mcginley_postselection-free_2023,Agrawal_Observing_2024,kamakari_experiment_2024}, and see refs.~\cite{Potter_2022,Fisher_2023} for reviews), and the MIPT has been shown to have implications for quantum error correction~\cite{Choi_2020,Fan_Self-organized_2021,Gullans_2020,Li_Statistical_2021,Fidkowski_how_2021,li_entanglement_2023}, quantum state tomography~\cite{Ippoliti_Learnability_2024,Akhtar_Measurement_2024}, and the computational difficulty of classically simulating a quantum system~\cite{Skinner_2019,Bao_Theory_2020}. 

So far, the primary approach for studying the MIPT is to cast the dynamics in the language of quantum circuits with mid-circuit measurements. In this context one can bring about the MIPT by varying either the frequency or the strength of measurements~\cite{Szyniszewski_Entanglement_2019,Szyniszewski_Universality_2020}.
Unlike other phase transitions in many-body systems, however, the MIPT has no signature in the expectation values of local operators or any other quantity that is linear in the system's density matrix. Instead, the MIPT is evident only in other nonlinear quantities associated with information theory, such as the entanglement entropy of a pure state or the purity of a mixed state that is produced by the dynamics~\cite{Skinner_2019,Gullans_2020}. 
At the level of theory or numerical simulations, one can straightforwardly study these quantities by examining the quantum state \emph{conditioned} on the outcomes of the mid-circuit measurements. But this conditioned state is difficult to study experimentally due to a key challenge known as the \textit{postselection} problem. Briefly, experimentally inferring a quantity that is nonlinear in the density matrix requires one to produce the same final state many times, but the final state depends on the history of measurement outcomes over the course of the dynamics. Since these measurement outcomes are random, one generally requires $\mathcal{O}(2^m)$ realizations of a given dynamics to obtain the same density matrix, where $m$ is the number of mid-circuit measurement outcomes~\cite{koh2023measurement}. Such an overhead becomes intractable if $m$ is a polynomial of the number of qubits in the system, which is usually the case. Some experiments~\cite{Noel_2022} simply absorb the postselection overhead at the cost of being unable to scale to larger system sizes, while others specialize to setups which cannot extend to generic unitary evolution through Haar-random gates~\cite{Ippoliti_postselection_2021,Li_Cross_2022, cote2022qubit,Hoke_Measurement-induced_2023}. 

Recent work has pointed out ways to avoid the postselection problem by constructing a quantity that incorporates the set of mid-circuit measurement outcomes for a given circuit realization into a classical calculation. For example, one can entangle one or more ancillary qubits (the \emph{probe}) with the system and study how the bipartite entanglement entropy between the probe and the system changes throughout the monitored dynamics~\cite{Gullans_2020}. In the entangling phase, the entanglement with the probe persists to a timescale that is exponential in the system size $\mathcal{N}$, while in the disentangling phase the time scale is $\mathcal{O}(\log \mathcal{N})$. But inferring the system-probe entanglement requires a classical decoder that can infer the reduced density matrix of the probe for a given measurement record (and a given set of unitary operators), and the difficulty of this classical decoding process scales exponentially with the system size for the most generic unitary dynamics. For this reason, existing experimental implementations~\cite{Noel_2022, kamakari_experiment_2024} of this probe-qubit protocol use only Clifford gates to evolve the system dynamically, since these admit a compact representation that is easy to simulate classically. Another approach based on the cross-entropy of measurement outcomes between identical circuits with differing initial conditions can also reveal the MIPT but suffers from the same exponential overhead of the classical decoding process for generic dynamics~\cite{garratt2023probing,Li_Cross_2022}.

Thus, in general an experimental study of the MIPT requires a tradeoff: since inferring the MIPT for the most general dynamics appears intractable with current algorithms, one must either remain restricted to small system sizes or sacrifice some generality of the dynamics to circumvent the exponential overhead of either the postselection problem or the classical decoding process. How one makes this sacrifice is a primary consideration for designing an experiment to study the MIPT.

Here we demonstrate a different approach to this tradeoff, which allows us to achieve the MIPT for generic, Haar-random unitary evolution without any postselection and using only a classical calculation whose complexity scales linearly with the number of qubits.
The key idea is to consider the setting of \emph{quantum trees}~\cite{Lopez-Piqueres_Mean-field_2020, nahum_measurement_2021,feng_measurement_2023,Feng_charge_2024, Sommers_dynamically_2024}, in which the system evolves through a tree-shaped tensor network of unitary operators and weak measurements. In this setting one can exploit the recursive structure of the tree to create a simple process for classical decoding. 
Further, the recursive structure enables an exact analytical solution for the critical point and the critical behavior~\cite{nahum_measurement_2021,feng_measurement_2023,Feng_charge_2024}, as well as numerical simulations of the relevant dynamics for large system sizes using the so-called pool method ~\cite{Miller_Weak-disorder_1994,Monthus_Anderson_2009,Garcia-Mata_Scaling_2017,Shi_Classical_2006,Murg_Simulating_2010,Silvi_Homogeneous_2010,Tagliacozzo_Simulation_2009,Li_Efficient_2012}.
In this paper we present a protocol for realizing the MIPT on a quantum tree and we implement our protocol experimentally on Quantinuum's H1-1 trapped-ion quantum computer~\cite{quantinuum2022}.
We obtain results for trees with up to four layers, which we show to be in good agreement with an exact analytical theory that describes the MIPT directly in our experiment. Our work therefore closes the gap between analytical theory and postselection-free observation of the MIPT on generic quantum trees.

\section{Model}
\label{sec:tree_model}

\begin{figure*}[htb!]
    \centering
    \includegraphics[width=1\linewidth]{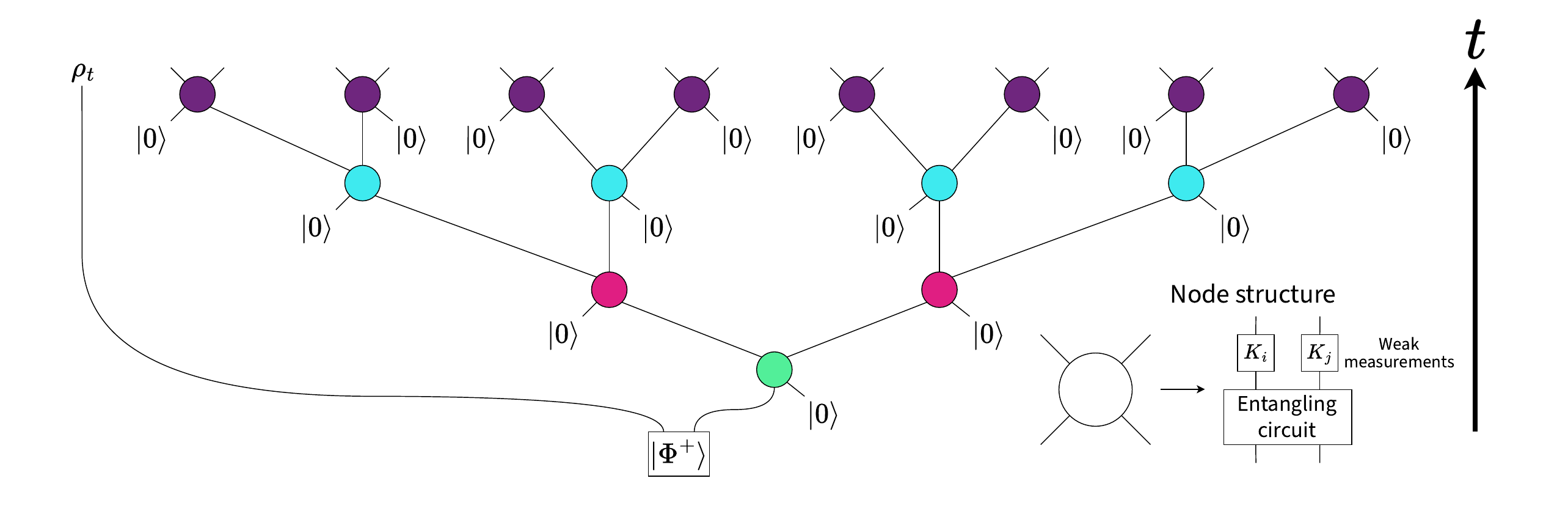}
    \caption{\justifying Diagram of the expansion process up to $t = 4$. Time flows from bottom to top. Each time step of the tree corresponds to a different color: $t = 1$ (green), $t = 2$ (pink), $t = 3$ (blue), and $t = 4$ (violet). At the root node (green), we input the first qubit from the Bell state $\ket{\Phi^+} \equiv \frac{1}{\sqrt{2}}\left(\ket{0}_P\ket{0}_R + \ket{1}_P \ket{1}_R\right)$, where the register $R$ holds the root qubit and the register $P$ holds the probe qubit. The latter acts as a probe for the phase transition, and is in state $\rho_t$ by the end of the evolution. As shown in the bottom-right of the figure, each node of the tree corresponds to first applying the entangling circuit in Fig.~\ref{fig:entangling-circuit} (with randomly chosen single-qubit unitaries), followed by the weak measurement circuit in Fig.~\ref{fig:weak-measurement-circuit} (which we depict using the Kraus operators $K_i$ and $K_j$), yielding measurement outcomes which we store. To limit clutter, the diagram does \emph{not} show the ancillary qubits used for the weak measurement circuits. The diagram also suppresses $\rho_t$'s dependence on the weak measurement strength $\theta$, the unitaries $\mathcal{U}$ encoding the circuit, and the weak measurement outcomes $M_w$.}
    \label{fig:quantum-tree-diagram}
\end{figure*}

\begin{figure*}[htb!]
    \justifying
    \includegraphics[width=1\linewidth]{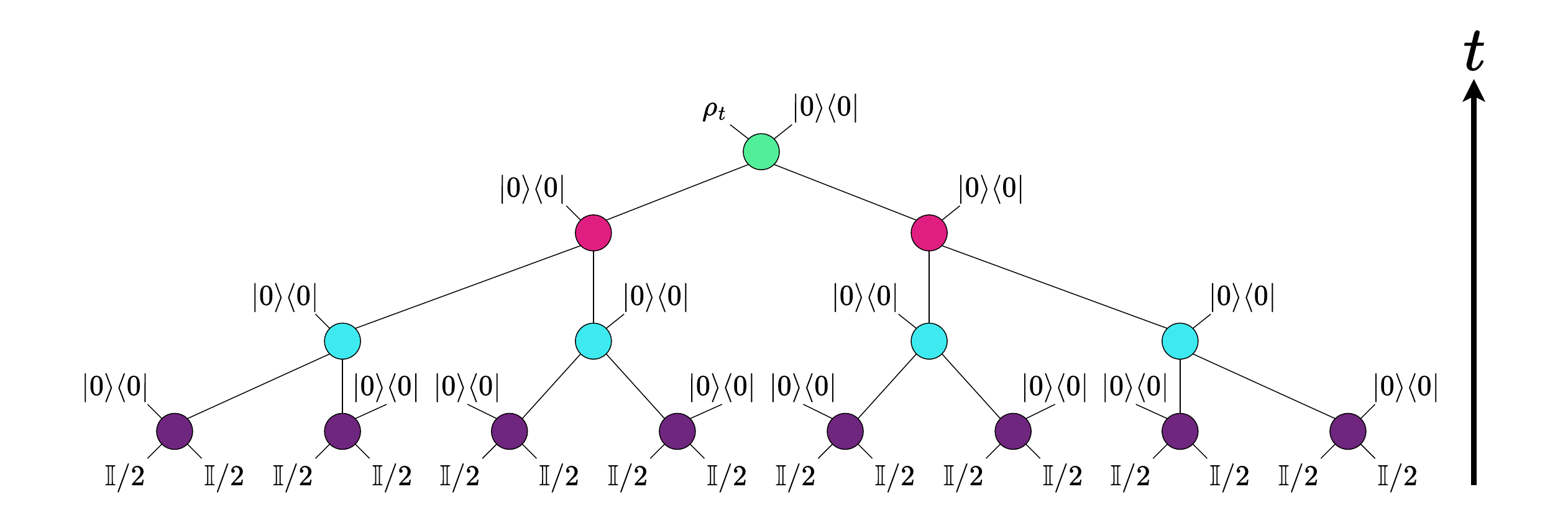}
    \caption{\justifying The collapse process for the dynamical quantum tree model for $t = 4$, yielding state $\rho_t$ (which is identical to $\rho_t$ in Fig.~\ref{fig:quantum-tree-diagram} given the same $\mathcal{U}$ and $M_w$). Note that we have suppressed $\rho_t$'s dependence on $\theta$, $\mathcal{U}$, and $M_w$. The tree is vertically reflected (about the $x$-axis) relative to Fig.~\ref{fig:quantum-tree-diagram}. To obtain the node structure in the collapse process, we reverse the order of time in the circuit of Fig.~\ref{fig:decomposed-node} (going right to left), and we take the complex conjugate of every circuit component. While in Fig.~\ref{fig:decomposed-node} we input an ancillary qubit in state $\ket{0} \bra{0}$, in the corresponding collapse process we project one of the output states to $\ket{0}\bra{0}$. The input state to each leaf node is $\rho_0 = \mathbb{I}/2$, where $\mathbb{I}$ is the $2\times 2$ identity matrix. Therefore, the full input state at the leaf nodes is $\rho_0^{\otimes \mathcal{N}} = \left( \mathbb{I}/2 \right)^{\otimes 2^t} = \mathbb{I}_{2^t}/2^{2^t}$, where $\mathbb{I}_{2^t}$ is the $2^t \times 2^t$ identity matrix.}
    \label{fig:collapse-tree-diagram}
\end{figure*}

In this paper, we focus on a dynamical quantum tree model similar to the one introduced in Refs.~\cite{nahum_measurement_2021,feng_measurement_2023}. For these models one can understand the final state as the output of a quantum circuit with mid-circuit measurements and a binary tree-like structure. There are two common choices for the flow of time in this model, called the \emph{expansion process} (Fig.~\ref{fig:quantum-tree-diagram}) and the \emph{collapse process} (Fig.~\ref{fig:collapse-tree-diagram}). In the expansion process, time flows from the root of the tree to the leaves. In contrast, the collapse process has time flowing from the leaves of the tree to the root. For quantities we are interested in, we can establish a mapping between both choices~\cite{feng_measurement_2023}, since both processes correspond to the same ensemble of tensor networks, and this mapping allows us to use whichever process is most convenient. The expansion process is a natural fit for our experiment, as we describe below, whereas the collapse process facilitates analytical analysis and numerical simulation (see Section~\ref{sec:reverse-construction-bloch-vector} and Section~\ref{appendix:theory}).

\begin{figure*}[ht]
    \centering
    \begin{subcaptionblock}[t]{0.3\textwidth}
        \begin{quantikz}
            \lstick{$\rho_{i,q}$} & \gate[2]{U_{\mathrm{ent}}}\gategroup[2, steps=3, style={dashed, draw=none}, label style={label position=below, anchor=north, yshift=-0.2cm}]{} & \gate{K_{m_r}} & \rstick{$\rho_{i+1,r}$}\\
            \lstick{$\ket{0} \bra{0}$} && \gate{K_{m_s}} & \rstick{$\rho_{i+1,s}$}
        \end{quantikz}
        \caption{Node in the expansion process of the dynamical quantum tree}
        \label{fig:decomposed-node}
    \end{subcaptionblock}
    \hspace{0.1\textwidth}
    \begin{subcaptionblock}[t]{0.3\textwidth}
        \begin{quantikz}
            \lstick{$\rho$} & \gate{U_1}\gategroup[2, steps=3, style={dashed}, label style={label position=below, anchor=north, yshift=-0.2cm}]{$U_{\mathrm{ent}}$} & \ctrl{1} & \gate{U_3} &\\
            \lstick{$\ket{0} \bra{0}$} & \gate{U_2} & \targ{} & \gate{U_4} &
        \end{quantikz}
        \caption{Entangling circuit} \label{fig:entangling-circuit}
    \end{subcaptionblock}
    \\
    \vspace{0.7cm}
    \begin{subcaptionblock}{0.8\textwidth}
        \begin{quantikz}
            \lstick{$\rho$} &\gategroup[2, steps=5, style={dashed}, label style={label position=below, anchor=north, yshift=-0.2cm}]{$K_{m}$} &&\ghost{H} & \ctrl{1} &&\\
            &\wireoverride{n}&\wireoverride{n}\lstick{$\ket{0}$} & \gate{R_Y (\theta)} & \targ{} & \meter{m}
        \end{quantikz}
        \,\,\, = \,\,\,
        \begin{quantikz}
            \lstick{$\rho$} &\gategroup[2, steps=5, style={dashed}, label style={label position=below, anchor=north, yshift=-0.2cm}]{$K_{m}$} && \gate[2]{ZZ(\phi)} &&&\\
            &\wireoverride{n} &\wireoverride{n}\lstick{$\ket{+}$} && \gate{R_X (\frac{\pi}{2})} & \meter{m}
        \end{quantikz}
        \caption{Weak measurement circuits} \label{fig:weak-measurement-circuit}
    \end{subcaptionblock}
    \caption{\justifying Decomposition of a node in Fig.~\ref{fig:quantum-tree-diagram} from time $i$ to $i+1$ in the expansion process of the dynamical quantum tree model. (a) Schematic circuit for a node, where the input states are $\rho_{i,q}$ and $\ket{0}\bra{0}$, followed by an entangling circuit $U_{\mathrm{ent}}$ and then two weak measurements, which we denote using the Kraus operators $K_{m_r}$ and $K_{m_s}$ of equation~\eqref{eq:kraus-operator}. (b) The entangling circuit to implement $U_{\mathrm{ent}}$. We randomly draw the unitaries $U_1$, $U_2$, $U_3$, and $U_4$ from the Haar measure. The CNOT gate generates entanglement between the qubits in the two registers. In general, the qubits in the dynamical quantum tree are entangled, so $\rho$ can be thought of as the reduced density matrix of a specific qubit. The bottom register contains an ancillary qubit that becomes entangled with the top register. Both qubits are then used for the next time step in the expansion process. (c) Two weak measurement circuits to implement the Kraus operator $K_m$ in equation~\eqref{eq:kraus-operator} with measurement strength $\theta$ and $\phi \equiv \pi/2 - \theta$. Both circuits require an ancillary qubit. The left circuit is the standard one for performing the weak measurement. Ignoring a global phase, the right circuit is logically equivalent (Section~\ref{sec:weak-measurement-circuits}) to the left one but uses the native two-qubit gate for the Quantinuum H1-1, defined as $ZZ(\phi) \equiv \exp{\left(-i(Z \otimes Z) \phi / 2\right)}$.}
    \label{fig:node-deconstruction}
\end{figure*}

In the expansion process, we begin with a Bell pair between a \emph{root} qubit and an ancillary \emph{probe} qubit. The latter qubit acts as a probe of the MIPT in the dynamical quantum tree model. We consider the layers of the tree to be discrete time steps: the root of the tree is at $t = 1$, the neighbors of the root node occur at $t = 2$, and so on. In Fig.~\ref{fig:quantum-tree-diagram}, we show that the number of qubits that are part of the system scales as $\mathcal{N} \equiv 2^t$ (excluding the probe qubit). Specifically, each node of the tree takes two inputs: a qubit from the previous time step and a fresh ancillary qubit in the state $\ket{0}$. The node involves two operations. First, we apply an entangling circuit involving a CNOT gate flanked by four single-qubit Haar-random unitary gates, as shown in Fig.~\ref{fig:entangling-circuit}. We use this choice of entangling circuit for experimental convenience, but our model also works in the case that the entangling circuit is just a two-qubit Haar-random gate. We use $\mathcal{U}$ to denote the list of unitary gates defining the quantum tree circuit for the expansion process. As we show in Section~\ref{appendix:reverse_map}, the choice of Haar-random gates allows us to understand the phase transition analytically and numerically. Second, we apply a weak measurement circuit (Fig.~\ref{fig:weak-measurement-circuit}) to each qubit. We can describe such a weak measurement with the Kraus operators:
\begin{equation} \label{eq:kraus-operator}
    K_{m} = \sin\left(\theta / 2\right)\mathbb{I}+\left[\cos\left(\theta / 2\right)-\sin\left(\theta / 2\right)\right]\ket{m}\bra{m},
\end{equation}
where $\mathbb{I}$ is the $2 \times 2$ identity matrix, $m \in \left\{0,1\right\}$, and $\theta$ is the \emph{measurement strength}, which we tune between $[\pi/2,\pi]$. We call the limit $\theta \rightarrow \pi/2$ the `weakest' limit, since channel~\eqref{eq:kraus-operator} becomes the identity. In contrast, the limit $\theta \rightarrow \pi$ is the `strong' limit, since the Kraus operators become projective measurements. We use $m$ to label the measurement outcome whose probability obeys Born's rule. In our model, the measurement strength $\theta$ is the same for all weak measurements. We then define $M_w$ to be the list of all weak measurement outcomes for the circuit.

At time $t$ of the dynamical quantum tree circuit, we have a pure state defined on $2^t+1$ qubits (ignoring the ancillary qubits that are used to implement the weak measurements using the right circuit in Fig.~\ref{fig:weak-measurement-circuit}). By tracing out every qubit but the probe qubit, we get the probe qubit's density matrix $\rho_t(\theta, \mathcal{U}, M_w)$, where we make explicit the dependence on time, measurement strength, the list of unitaries $\mathcal{U}$, and the weak measurement outcomes $M_w$. We are interested in how $\rho_t(\theta, \mathcal{U}, M_w)$ evolves over time for different measurement strengths.

Previous work~\cite{nahum_measurement_2021,feng_measurement_2023} considered a slight variant on our setup: instead of the entangling circuit in Fig.~\ref{fig:entangling-circuit}, one samples a two-qubit Haar-random unitary. That work showed that as $t \to \infty$, the probe qubit undergoes a phase transition from a mixed phase ($\Tr{\left[\rho^2_{t\rightarrow \infty}(\theta, \mathcal{U}, M_w)\right]} < 1$) when the measurement strength $\theta$ is less than a critical threshold $\theta_c$ to a pure phase ($\Tr{\left[\rho^2_{t\rightarrow \infty}(\theta, \mathcal{U}, M_w)\right]} = 1$ with probability approaching $1$) when $\theta > \theta_c$. To offer another perspective on the two phases, consider the bipartite entanglement entropy $athcal{S}_{P}$ between the probe qubit in register $P$ and the qubits of the system in register $S$. When $\theta<\theta_c$, the state of the probe qubit is mixed, so $\mathcal{S}_P$ remains finite even with $t\to\infty$. This tells us that entanglement remains between the probe qubit and the system in the thermodynamic limit $t\to\infty$. At the same time, $\mathcal{S}_P$ decays to $0$ with the increase of system size in the pure state phase $\theta>\theta_c$. Thus, the probe qubit becomes completely disentangled from the system in the thermodynamic limit $t\to\infty$.

\begin{figure}[ht]
    \centering
    \includegraphics[width=1\columnwidth]{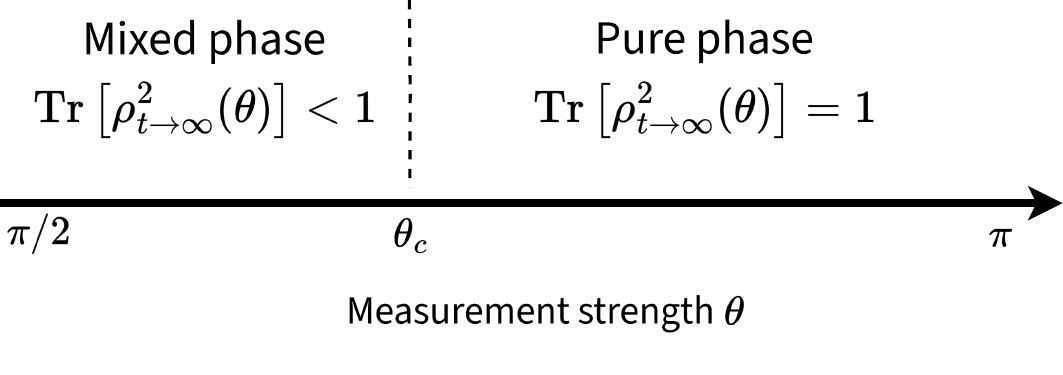}
    \caption{\justifying Phase diagram for the dynamical quantum tree model as a function of the measurement strength $\theta$. For $\theta < \theta_c \approx 2.2$, the probe qubit's state $\rho_{t\rightarrow \infty}(\theta)$ is mixed, while the state is pure for $\theta > \theta_c$. Here, we suppress the state's dependence on $\mathcal{U}$ and $M_w$.}
    \label{fig:phase-diagram}
\end{figure}

A similar phase transition occurs in our setup, with only the critical measurement strength $\theta_c$ changing. (Since $\theta_c$ is defined differently in Ref.~\cite{feng_measurement_2023}, we need to multiply that value by $2$ to compare with the result in this work.) We sketch the phase diagram in Fig.~\ref{fig:phase-diagram}. To characterize the phase transition, we examine the smallest eigenvalue $Z_t(\theta, \mathcal{U}, M_w)$ of the probe qubit's state for finite $t$:
\begin{equation} \label{eq:rho_k}
    \rho_t(\theta, \mathcal{U}, M_w)=
    \nu\begin{pmatrix}
        1-Z_t(\theta, \mathcal{U}, M_w)&0\\
        0&Z_t(\theta, \mathcal{U}, M_w)
    \end{pmatrix}\nu^{\dagger},
\end{equation}
where $\nu$ is the eigenbasis (we suppress its dependence on $t$, $\theta$, $\mathcal{U}$, and $M_w$ for brevity). Since $\theta$ defines the phases of the system, the typical value $Z^{\mathrm{typ}}_t(\theta)$ acts as an order parameter of the phase transition, which we get by computing the geometric mean of $Z_t(\theta,\mathcal{U},M_{w})$~\cite{derrida_survival_2007,nahum_measurement_2021,feng_measurement_2023}:
\begin{equation} \label{eq:typical-Z}
    \ln Z^{\mathrm{typ}}_t(\theta) \equiv \mathbb{E}_{\mathcal{U}, M_w} \left[ \ln Z_t(\theta,\mathcal{U},M_w)\right].
\end{equation}
The notation $\mathbb{E}_{\mathcal{U}, M_w} \left[ \cdots\right]$ indicates we average over the ensemble of independently sampled single-qubit Haar-random unitaries and measurement outcomes. Note that equation~\eqref{eq:typical-Z} is only defined for $Z_t(\theta,\mathcal{U},M_w) > 0$. Therefore, we exclude any cases where $Z_t(\theta,\mathcal{U},M_w) = 0$ when taking the average. Observe that $\theta = \pi \implies Z_t(\theta,\mathcal{U},M_w)=0$ for any $\mathcal{U}$ and $M_w$ since the weak measurements become projective, so we define $Z_t^{\mathrm{typ}}(\theta) \equiv 0$. In Sections~\ref{appendix:reverse_map} and~\ref{appendix:theory}, we discuss the order parameter further and explain that we can obtain an analytical solution of the phase transition in our dynamical quantum tree model by mapping the dynamics to a Fisher-KPP type of equation~\cite{Derrida_Polymers_1988,derrida_survival_2007,nahum_measurement_2021,feng_measurement_2023,Feng_charge_2024}. Using this mapping, we find the critical measurement strength is $\theta_c=2.2142(2)$. Furthermore, due to the recursive structure of the dynamical quantum tree model, there exists a numerical method to easily simulate the dynamics of $\rho_t(\theta,\mathcal{U},M_w)$ for large enough system sizes~\cite{Miller_Weak-disorder_1994,Monthus_Anderson_2009,Garcia-Mata_Scaling_2017,Shi_Classical_2006,Murg_Simulating_2010,Silvi_Homogeneous_2010,Tagliacozzo_Simulation_2009,Li_Efficient_2012} (we provide details in Section~\ref{appendix:pool_method} and simulate the dynamics up to $t = 800$). There also exists a method which can efficiently estimate an order parameter related to equation~\ref{eq:typical-Z} given only the sampled unitaries, the weak measurement outcomes and a measurement of the probe qubit. The latter method is the basis of our experimental protocol, which we discuss in Section~\ref{sec:methods}.

Before we end this section, we want to emphasize that our phase transition is a property of the dynamical quantum tree model, not the probe qubit. In other words, the probe qubit is not necessary for this phase transition to occur. While the value of $Z_t(\theta,\mathcal{U},M_w)$ is a measure of the purity of the probe qubit, we can also characterize the phase transition by our ability to predict the initial state using the weak measurements inside the circuit. We show below that the latter does not require a probe qubit. Therefore, while we refer to the state of the probe qubit for convenience, in our experiment we use the aforementioned predictability to detect the phase transition. In Section~\ref{sec:probe}, we give a postselection-free efficient protocol to probe the phase transition, which quantifies the predictability of the root qubit's initial state ($\ket{0}$ or $\ket{1}$) given the mid-circuit weak measurement outcomes.

\section{Results} \label{sec:results}

While the quantity $Z^{\mathrm{typ}}_t(\theta)$ in equation~\eqref{eq:typical-Z} characterizes the MIPT in the expansion process of the dynamical quantum tree model, it is not an ideal quantity to estimate in experiment. The reason is that determining $Z_t(\theta, \mathcal{U}, M_w)$ for a given $t$, $\theta$, and $\mathcal{U}$ requires multiple realizations to get the same measurement outcomes $M_w$, a manifestation of the post-selection problem. Our workaround is to consider the simpler quantity:
\begin{equation} \label{eq:averaged}
    Z_t (\theta) \equiv \mathbb{E}_{\mathcal{U},M_w} \left[ Z_t(\theta,\mathcal{U},M_w) \right].
\end{equation}
At first, $Z_t(\theta)$ does not appear any easier to determine than $Z^{\mathrm{typ}}_t(\theta)$. However, averaging over a linear quantity in $Z_t(\theta, \mathcal{U}, M_w)$ does not require us to explicitly evaluate $Z_t(\theta, \mathcal{U}, M_w)$, and we show in Section~\ref{sec:probe} that this allows us to bypass the postselection problem. Furthermore, previous work showed that $Z_{t\to\infty} (\theta) \sim Z_{t\to\infty}^{\mathrm{typ}}(\theta)$~\cite{feng_measurement_2023} when close to the critical point $\theta_c$. We obtain a similar result in Section~\ref{appendix:theory}. This motivates us to use $Z_t(\theta)$ as the order parameter instead. In Sections~\ref{sec:probe} and~\ref{sec:reverse-construction-bloch-vector}, we show how to construct an estimate $\hat{Z}_{t,N_c}(\theta)$, averaged over some number $N_c$ of random instances that we encode as quantum circuits and some number $N_s$ of repetitions (shots) per circuit. We further discuss how to construct these circuits in Section~\ref{sec:circuit-construction} and our experimental resources in Section~\ref{sec:experimental-details}.

\begin{figure}
    \centering
    \includegraphics[width=1\columnwidth]{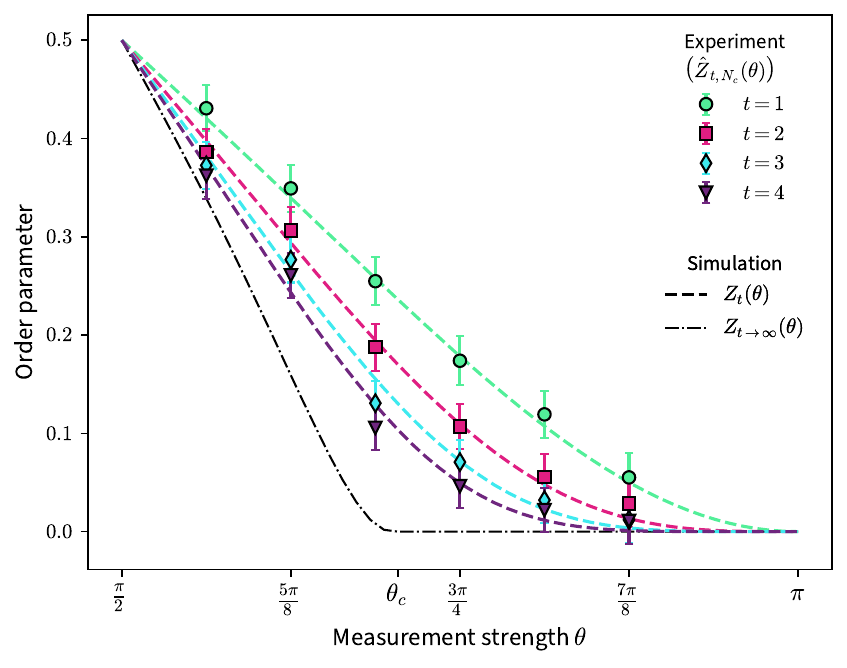}
    \caption{\justifying Experimental results for the order parameter $\hat{Z}_{t,N_c}(\theta)$ on the Quantinuum H1-1. We construct each data point for $t = 4$ as in Algorithm~\ref{alg:estimate_Z}, consisting of an average over $N_c = 834$ random circuits, with $N_s = 8$ shots per circuit. Error bars show a $95\%$ confidence interval and have size $1.96 \cdot E_{t,N_c}(\theta)$, where $E_{t,N_c}(\theta)$ is defined in line~\ref{line:error} of Algorithm~\ref{alg:estimate_Z}, as explained in Section~\ref{sec:probe}. We derive the $t \leq 3$ data points from the $t = 4$ data by truncating the measurement record (see Section~\ref{sec:experimental-details}). Dashed lines are the simulated predictions of $Z_t(\theta)$ using the pool method of Section~\ref{appendix:pool_method} with pool size $10^6$. The dot-dashed line indicates the simulated prediction for $Z_{t \rightarrow \infty}(\theta)$ using the pool method (in the figure, $t = 800$ which has converged). All simulated results (dashed and dot-dashed lines) have an error bar size given by $1.96 \cdot \mathrm{SE}$, where $\mathrm{SE}$ is the standard error of the mean for that data point over the $10^6$ values in the corresponding pool. All error bars for the simulated values are less than $2.1 \cdot 10^{-4}$ (see Section~\ref{appendix:pool_method}), so we omit them to reduce clutter.}
    \label{fig:experimental-results}
\end{figure}

In Fig.~\ref{fig:experimental-results}, we present the experimental results for the order parameter $\hat{Z}_{t,N_c}(\theta)$ of the dynamical quantum tree model up to $t = 4$, where $N_c = 834$ and $N_s = 8$. We note several observations. First, the experimental data points follow the the simulated theory curves (dashed lines), suggesting that the experimental protocol can indeed estimate the order parameter and that noise in the Quantinuum H1-1 does not significantly affect the results. Second, for a fixed $t$, $\hat{Z}_{t,N_c}(\theta)$ decreases as the measurement strength $\theta$ increases. Recall that increasing the measurement strength corresponds to weak measurements that approach projective measurements. Intuitively, projective measurements will increase the purity of the probe qubit, leading to $\hat{Z}_{t,N_c}(\theta) \rightarrow 0$. Third, for a fixed $\theta$, the value of $\hat{Z}_{t,N_c}(\theta)$ approaches the asymptotic result $Z_{t \rightarrow \infty}(\theta)$ as the tree size $t$ is increased.

Our results indicate signatures of the phase transition associated with our dynamical quantum tree model. In particular, if one focuses on a specific $t$, there are two qualitative regimes: a regime where the order parameter remains positive, followed by a regime where the order parameter decays quickly to zero. As $t \rightarrow \infty$, the measurement strength that separates these two regimes approaches $\theta_c$ from the right. Such a behavior suggests that as $t \rightarrow \infty$, the state of a probe qubit initially entangled with the root qubit remains mixed for $\theta < \theta_c$, since $Z_{t\rightarrow \infty}(\theta) > 0$. In contrast, for any measurement strength $\theta > \theta_c$, the probe qubit becomes pure.

\section{Discussion}
\label{sec:discussion}

In this work we have reported experimental signatures of the MIPT in a dynamical quantum tree. Our primary result is Fig.~\ref{fig:experimental-results}, which indicates a purification transition as a function of increasing measurement strength realized on the Quantinuum H1-1 quantum computer. The experimental data shows excellent agreement with theoretical results without any error mitigation. 

Our ability to realize the MIPT without the usual problems of postselection or an exponentially difficult classical decoding process arises because of the structure of the quantum tree. The recursive structure of the tree not only makes decoding process easy (with a complexity that scales with the number of qubits), but it enables a theoretically tractable solution for the properties of the MIPT (as pointed out in~\cite{nahum_measurement_2021, feng_measurement_2023}). Our results represent the first realization, to our knowledge, of an MIPT with Haar-random unitary evolution and without postselection.

One key ingredient to our analysis is that expectation values of any function of the probe qubit density matrix are invariant to any single-qubit rotation. We make use of this property in equation~\eqref{eq:mean-X_t} and in Section~\ref{appendix:reverse_map}. In particular, this property enables the recursive `decoding' technique by which we can reconstruct the probe qubit density matrix using the measurement record. Consequently, our experimental realization is limited primarily only by the available number of qubits in the quantum computer. Without such an efficient technique, the postselection problem would limit our ability to probe the phase transition at larger system sizes. It is therefore interesting to ask whether efficient decoding is a property of a broader class of tree models.

We note also that recent work shows that the dynamical quantum trees with Abelian/non-Abelian symmetries can be understood theoretically~\cite{Feng_charge_2024}. In such systems, a `sharpening' transition emerges in addition to the usual MIPT~\cite{agrawal_entanglement_2022,majidy_critical_2023,Feng_charge_2024,Agrawal_Observing_2024}. Experimental realization of other quantum tree models, especially those with non-Abelian symmetry, can be an interesting direction for future work. 
\section{Methods} \label{sec:methods}

\subsection{Experimental probe of the phase transition} \label{sec:probe}

The main experimental challenge of detecting a measurement-induced phase transition is the postselection problem: since $\rho_t(\theta,\mathcal{U},M_w)$ in equation~\eqref{eq:rho_k} depends on $M_w$, which follows Born's rule, one requires $\mathcal{O}\left(2^{\lvert M_w \rvert}\right) = \mathcal{O}\left(2^{\mathcal{N}}\right) = \mathcal{O}(2^{2^t})$ many circuit realizations to determine $Z_t(\theta,\mathcal{U},M_w)$ through techniques such as quantum state tomography~\cite{Banaszek_focus_2013}. While several experiments have reported evidence of other measurement-induced phase transitions~\cite{Noel_2022,koh2023measurement,Hoke_Measurement-induced_2023,kamakari_experiment_2024,Agrawal_Observing_2024}, the protocols are often resource-intensive~\cite{koh2023measurement} and do not efficiently scale with system size~\cite{Noel_2022,Hoke_Measurement-induced_2023}. In this section, we describe an efficient experimental protocol from ref.~\cite{feng_measurement_2023} that avoids the postselection problem via a classical decoding algorithm whose complexity is linear in the system size, allowing us to observe signatures of the measurement-induced phase transition on the Quantinuum H1-1.

To begin, we note that although the phase transition is defined for $t \rightarrow \infty$, we can only implement a quantum circuit for finite $t$. Therefore, whenever we discuss the state $\rho_t(\theta, \mathcal{U}, M_w)$, we have in mind the state of the probe qubit evolved for $t$ time steps of the dynamical quantum tree. We can write the state as:
\begin{equation} \label{eq:bloch-state}
    \rho_t(\theta, \mathcal{U}, M_w) = \frac{1}{2}\left[\mathbb{I}+\vec{\sigma}\cdot\vec{n}_t(\theta, \mathcal{U}, M_w)\right],
\end{equation}
where $\vec{\sigma}=(\sigma_x,\sigma_y,\sigma_z)$ is the Pauli vector consisting of the Pauli matrices $\sigma_x$, $\sigma_y$, and $\sigma_z$, and $\vec{n}_t(\theta, \mathcal{U}, M_w)$ is the Bloch vector. By computing the eigenvalues of $\rho_t(\theta, \mathcal{U}, M_w)$ in equation~\eqref{eq:bloch-state} and comparing with equation~\eqref{eq:rho_k}, we find $|\vec{n}_t(\theta, \mathcal{U}, M_w)| = 1-2Z_t(\theta, \mathcal{U}, M_w)$. With $\mathcal{U}$ and $M_w$, we can calculate $\rho_t(\theta, \mathcal{U}, M_w)$ (see Section~\ref{sec:reverse-construction-bloch-vector}). However, as we mentioned in Section~\ref{sec:results}, it is not easy to calculate the average in equation~\eqref{eq:typical-Z} since determining $Z_t(\theta,\mathcal{U},M_w)$ requires multiple copies of $\rho_t(\theta, \mathcal{U}, M_w)$ and therefore the same measurement outcomes $M_w$, which generally leads to the postselection problem. We therefore consider $Z_t(\theta)$ in equation~\eqref{eq:averaged} as the order parameter, which retains the same behavior of $Z^{\mathrm{typ}}_t(\theta)$ near the MIPT but is easier to extract using our protocol below.

In Algorithm~\ref{alg:estimate_Z}, we describe an experimental protocol to probe the phase transition by estimating $Z_t(\theta)$ for various $t$ and $\theta$. The logic of Algorithm~\ref{alg:estimate_Z} is as follows. We sample $N_c$ random circuits encoding instances of the dynamical quantum tree model, described by the parameters $t$, $\theta$, and $\mathcal{U}$ from line~\ref{line:sample-unitaries}. We run each circuit $N_s$ times, with each repetition (shot) yielding a measurement record $\mathcal{M} = (m_0, m_1, \ldots)$ (line~\ref{line:sample-circuit}) encoding the measurement outcomes for the probe qubit ($m_0$) and the weak measurements $M_w = (m_1, m_2, \ldots)$. We then compute a classical prediction for the Bloch vector (line~\ref{line:classical-prediction}), which allows us to calculate the quantity $X_t$ (line~\ref{line:X_i}), similar to the method in ref.~\cite{Hoke_Measurement-induced_2023}. For a given circuit, we average over the $N_s$ shots to get $\bar{X}_t$ (line~\ref{line:bar_X}). Averaging again over the $N_c$ circuits yields an estimate $\hat{Z}_{t,N_c}(\theta)$ (line~\ref{line:estimator}) for $Z_t(\theta)$.

To prove that $\hat{Z}_{t,N_c}(\theta)$ is a reasonable estimate, we analyze the expectation value of $X_t$ over the ensemble of unitaries $\mathcal{U}$ and the measurement outcomes $\mathcal{M}$. We can focus on the expectation value of $X_t$ since each observation is weighted equally in the average on line~\ref{line:estimator} of Algorithm~\ref{alg:estimate_Z}. We find:
\begin{align} \label{eq:mean-X_t}
    \mathbb{E}_{\mathcal{U}, \mathcal{M}} \left[ X_t \right] &= \frac{1}{2} - \mathbb{E}_{\mathcal{U}, M_w, m_0} \left[ \frac{(-1)^{m_0}}{\mathrm{sign}\left[n_{t}^z(\theta, \mathcal{U}, M_w)\right]} \right] \notag \\
    &= \frac{1}{2} - \mathbb{E}_{\mathcal{U}, M_w} \left[ \frac{\mathbb{E}_{m_0|\mathcal{U}, M_w} \left[ (-1)^{m_0} \vert \mathcal{U}, M_w \right]}{\mathrm{sign}\left[n_{t}^z(\theta, \mathcal{U}, M_w)\right]} \right]  \notag \\
    &= \frac{1}{2} - \mathbb{E}_{\mathcal{U}, M_w} \left[\frac{n_t^z(\theta, \mathcal{U}, M_w)}{\mathrm{sign}\left[n_{t}^z(\theta, \mathcal{U}, M_w)\right]} \right] \notag \\
    &= \frac{1}{2} - \mathbb{E}_{\mathcal{U}, M_w} \left[ \lvert n_t^z(\theta, \mathcal{U}, M_w) \rvert \right] \notag \\
    &= \frac{1}{2} - \mathbb{E}_{\vec{n}_t \equiv \vec{n}_t(\theta, \mathcal{U}, M_w)} \left[ \lvert n_t^z \rvert \right] \notag \\
    &= \frac{1}{2} - \mathbb{E}_{\lvert \vec{n}_t\rvert} \left[ \lvert \vec{n}_t \rvert \cdot \mathbb{E}_{\hat{n}_t \,\, \vert \,\, \lvert \vec{n}_t\rvert} \left[ \lvert \hat{n}_t \cdot \hat{z} \rvert \,\, \vert \,\, \lvert \vec{n}_t\rvert  \right] \right] \notag \\ 
    &= \frac{1}{2} - \mathbb{E}_{\lvert \vec{n}_t \rvert} \left[ \vert \vec{n}_t \rvert \right] \cdot \mathbb{E}_{\hat{n}_t} \left[ \lvert \hat{n}_t \cdot \hat{z} \rvert \right] \notag \\
    &=\frac{1}{2} - \frac{1}{2} \mathbb{E}_{\lvert \vec{n}_t \rvert}\left[ \lvert \vec{n}_t \rvert \right] \notag \\
    &= Z_t(\theta)
\end{align}
The first equality is by definition of the expectation value. The second equality evaluates the conditional expectation of $(-1)^{m_0}$ given $\mathcal{U}$ and $M_w$ using the law of total expectation: $\mathbb{E}_{X,Y}\left[f(X,Y)\right] = \mathbb{E}_{X}\left[ \mathbb{E}_{Y|X} \left[f(X,Y)\right] \right]$. To get the third equality, use equation~\eqref{eq:bloch-state} to find $\Pr[m_0 \vert \mathcal{U}, M_w] = \frac{1}{2} \left[ 1 + (-1)^{m_0} n_t^z(\theta, \mathcal{U}, M_w) \right]$ and directly compute the expectation value:
\begin{align} \label{eq:expectation-m0}
    \mathbb{E}_{m_0}\left[(-1)^{m_0} \vert \mathcal{U}, M_w\right] &= \sum_{m_0 \in \left\{0,1\right\}} \Pr[m_0 \vert \mathcal{U}, M_w] (-1)^{m_0} \notag \\ 
    &= \frac{1}{2} \sum_{x = \pm 1} x \left[ 1 + x n_t^z(\theta, \mathcal{U}, M_w) \right] \notag \\
    &= n_t^z(\theta, \mathcal{U}, M_w).
\end{align}
The fourth equality of equation~\eqref{eq:mean-X_t} comes from using $c = \lvert c \rvert \cdot \mathrm{sign}(c)$ for any real number $c$. The fifth equality simply rewrites the expectation over $\mathcal{U}$ and $M_w$ in terms of the expectation over the Bloch vector $\vec{n}_t (\theta, \mathcal{U}, M_w)$. The sixth equality splits the expectation over $\vec{n}_t(\theta,\mathcal{U}, M_w)$ into a part with its magnitude ($\lvert \vec{n}_t \rvert$) and a part with its projection to in the $\hat{z}$ direction (yielding $\lvert \hat{n}_t \cdot \hat{z} \rvert = \lvert \cos\vartheta \rvert$, where $\vartheta$ is the polar angle in spherical coordinates) using the law of total expectation again. The seventh equality uses the observation that the magnitude and direction of the Bloch vector are independent due to the Haar-invariance of the expectation value under our choice of $\mathcal{U}$. Furthermore, the Haar-invariance implies that $\hat{n}_t$ is uniformly distributed about the unit sphere, allowing us to explicitly evaluate the angular expectation for the eighth equality:
\begin{equation} \label{eq:angular-expectation}
    \mathbb{E}_{\hat{n}_t} \left[ \lvert \hat{n}_t \cdot \hat{z} \rvert \right] = \frac{1}{4\pi} \int_{0}^{2\pi} \mathrm{d}\phi \int_0^{\pi} \mathrm{d} \vartheta \lvert \cos\vartheta \rvert \sin\vartheta = \frac{1}{2}. 
\end{equation}
The final equality of equation~\eqref{eq:mean-X_t} comes from the relation $\lvert \vec{n}_t(\theta, \mathcal{U}, M_w) \rvert = 1 - 2 Z_t(\theta, \mathcal{U}, M_w)$ (see below equation~\eqref{eq:bloch-state}) and rewriting the expectation value to be over $\mathcal{U}$ and $M_w$ once again. Note that $\lim_{N_c \rightarrow \infty} \hat{Z}_{t,N_c}(\theta) = Z_t (\theta)$. Therefore, by averaging over a large number of circuits $N_c$ as in line~\ref{line:estimator} of Algorithm~\ref{alg:estimate_Z} and assuming the noise in the Quantinuum H1-1 is not too large, we can estimate $Z_t(\theta)$ using $\hat{Z}_{t,N_c}(\theta)$. We calculate the standard error $E_{t,N_c}(\theta)$ of the mean for $\hat{Z}_{t,N_c}(\theta)$ on line~\ref{line:error}, where each of the $N_c$ data points in $\mathcal{D}$ is independent.

There is no postselection problem in Algorithm~\ref{alg:estimate_Z} because the quantity on line~\ref{line:classical-prediction} is efficient (in $\mathcal{N}$) to compute for any circuit realization, as we explain in Section~\ref{sec:reverse-construction-bloch-vector}. Furthermore, while the probe qubit aided us in our analysis, no unitaries act on it throughout the evolution. As such, in terms of constructing the estimate $\hat{Z}_{t,N_c}(\theta)$, we can measure the probe qubit immediately after preparing the Bell state. This is equivalent to not even using a probe qubit and simply preparing the root qubit to be $\ket{0}$ or $\ket{1}$ with equal probability (rather than half of a Bell pair). We opt to do this in our experiment to reduce the qubit count by one.

One may then ask how to interpret $\hat{Z}_{t,N_c}(\theta)$ and the phase transition in the experiment if we do not use a probe qubit. As we mentioned at the end of Section~\ref{sec:tree_model}, the phase transition is a property of the dynamical quantum tree model itself, not the probe qubit. In equation~(\ref{eq:mean-X_t}), we see that averaging over $X_t$ gives $Z_t(\theta)$. Note that $(1/2 - X_t) \in \pm 1$ indicates agreement ($+1$) or disagreement ($-1$) between $m_0$ and our prediction of it using the weak measurement outcomes $M_w$ (line~\ref{line:X_i} of Algorithm~\ref{alg:estimate_Z}). While we can view $m_0$ as a projective measurement of the probe qubit after the evolution, because the initial state of the probe qubit and root qubit is $\ket{\Phi^+}$, we can equally view $m_0$ as a uniformly random choice of initial product state $\ket{m_0}_R$ for the root qubit. The probabilities associated to measuring $\mathcal{M}$ remain the same in both cases. This means that we can understand the phase transition through either the purification transition of probe qubit or a phase transition in our ability to predict the initial state using the mid-circuit measurement outcomes. In other words, one can view our experimental estimate $\hat{Z}_{t,N_c}(\theta)$ as quantifying the degree to which the dynamical quantum tree \emph{would} purify ($Z_t(\theta)\to 0$) a probe qubit entangled with the root qubit as in Fig.~\ref{fig:quantum-tree-diagram}.

\begin{algorithm}[hbt!]
\caption{Estimate $Z_t(\theta)$}\label{alg:estimate_Z}
\Input{Measurement strength $\theta$, time $t$, number of circuits $N_c$, and shots per circuit $N_s$.}
$\mathcal{D} \gets \left[ \,\,\,\, \right]$ \Comment*[r]{Initialize an empty list.}
\For{$i \in \left\{1,2, \ldots, N_c\right\}$}{
    $\mathcal{U} \gets \mathrm{SampleUnitaries}(4(2^t-1))$ \label{line:sample-unitaries} \Comment*[r]{Independently sample $4 (2^{t}-1)$ single-qubit Haar-random unitaries.}
    $d \gets \left[ \,\,\,\, \right]$\\
    \For{$j \in \left\{1, 2, \ldots, N_s\right\}$}{
        $\mathcal{M} \gets \mathrm{SampleCircuit}(t, \theta, \mathcal{U})$ \label{line:sample-circuit} \Comment*[r]{Measurement record $\mathcal{M} = (m_0, m_1, \ldots) \in \left\{ 0,1\right\}^\ast$, Probe qubit measurement outcome is $m_0$.}
        $M_w \gets \mathcal{M}[1:]$ \Comment*[r]{$M_w = (m_1,m_2, \ldots)$}
        $\vec{n}_t(\theta, \mathcal{U}, M_w) \gets \mathrm{ClassicalPrediction}(t, \theta, \mathcal{U}, M_w)$ \label{line:classical-prediction} \Comment*[r]{Use the method of Section~\ref{sec:reverse-construction-bloch-vector}.}
        $X_t \gets \frac{1}{2} - (-1)^{m_0} /  \mathrm{sign}\left[n_{t}^z(\theta, \mathcal{U}, M_w)\right]$ \label{line:X_i}\\
        $\mathrm{Append}(d, X_t)$\\
    }
    $\bar{X}_t \gets \frac{1}{N_s} \sum_{j = 1}^{N_s} d_i$ \label{line:bar_X}\\
    $\mathrm{Append}(\mathcal{D}, \bar{X}_t)$\\
}
$\hat{Z}_{t,N_c} (\theta) \gets \frac{1}{N_c} \sum_{i=1}^{N_c} \mathcal{D}_i$ \label{line:estimator}\\
    $E_{t,N_c}(\theta) \gets \sqrt{\frac{\sum_{i=1}^{N_c}\left[\mathcal{D}_i - \hat{Z}_{t,N_c}(\theta)\right]^2}{N_c(N_c-1)}}$ \label{line:error}\\
\Return{$\hat{Z}_{t,N_c}(\theta)$, $E_{t,N_c}(\theta)$}
\end{algorithm}

\subsection{Reverse construction of the Bloch vector}
\label{sec:reverse-construction-bloch-vector}

On line~\ref{line:classical-prediction} of Algorithm~\ref{alg:estimate_Z}, we claim one can efficiently (as as function of $\mathcal{N}$) compute a classical prediction for $\vec{n}_t$ given the unitaries $\mathcal{U}$ and the measurement outcomes $M_w$. To prove our claim, we first note that there are multiple ways to compute $\vec{n}_t$. One way is to build a $(2^{2^t}, 2)$ matrix $T_t \equiv T_t(\theta, \mathcal{U}, M_w)$ that maps the initial state $\ket{\Phi^+}$ of the probe and root qubits to a final non-normalized state on all $2^t+1$ qubits, as in Fig.~\ref{fig:quantum-tree-diagram}. One can interpret the tensor as containing the unitaries in $\mathcal{U}$ and the weak measurement Kraus operators of equation~\eqref{eq:kraus-operator} for each weak measurement in $M_w$. For clarity, we suppose $T_t$ only acts on the $R$ register and acts as the identity otherwise, so $\ket{\psi_t} \equiv \frac{1}{\sqrt{2}} \left(T_t\ket{0}_R \ket{0}_P + T_t\ket{1}_R \ket{1}_P\right)$. Taking a partial trace over the $2^t$ qubits of the system, we find the probe qubit's state: 
\begin{align} \label{eq:probe-state-expansion}
    \rho_t(\theta,\mathcal{U},M_w) &= \frac{\Tr_{\bar{P}}{\left[\ket{\psi_t} \bra{\psi_t}\right]}}{ \Tr{\left[\ket{\psi_t} \bra{\psi_t}\right]}} \notag \\
    &= \frac{\frac{1}{2} \sum_{i,j = 0}^1 \bra{i} T_t T_t^{\dagger}  \ket{j}_R \otimes \ket{j} \bra{i}_P}{\Tr{\left[\ket{\psi_t} \bra{\psi_t}\right]}} \notag \\
    &= \frac{\left(T_t^{\dagger}T_t\right)^{T}}{\Tr\left[T_t^{\dagger}T_t\right]},
\end{align}
where $\bar{P}$ refers to every qubit but the probe qubit. The first equality is the definition of the probe qubit state, the second equality substitutes the definition of $\ket{\psi}_t$ and evaluates the trace in the numerator. The third equality recognizes the inner products on the second line as matrix elements of $\left(T_t^{\dagger}T_t\right)^{T}$, and uses $\Tr{\left[\ket{\psi_t} \bra{\psi_t}\right]} = \frac{1}{2} \Tr\left[T_t^{\dagger}T_t \right]$.

Notice that we can insert a rescaled $2^t \times 2^t$ identity matrix $\mathbb{I}_{2^t} / 2^t$ between the two tensors in the numerator and denominator of equation~\eqref{eq:probe-state-expansion} without changing the state. Yet, such a substitution admits a new interpretation of the equation and thus a way for computing $\vec{n}_t$. Namely, we can view $\rho_t(\theta, \mathcal{U}, M_w)$ as in the collapse process of Fig.~\ref{fig:collapse-tree-diagram}, where we evolve an initial maximally mixed state $\mathbb{I}_{2^t} / 2^t$ by the matrix $T_t^T$. Unfortunately, constructing $T_t$ is inefficient in $t$. This prohibits us from computing $\rho_t$ by building the non-normalized state $\ket{\psi_t}$ in the expansion process and then taking the partial trace. However, it is possible to compute the final state (and its Bloch vector $\vec{n}_t$) in the collapse process using an iterative method that does not require $T_t$ and is therefore efficient in the system size $\mathcal{N} = 2^t$.

Consider Fig.~\ref{fig:collapse-tree-diagram} for a given $\mathcal{U}$ and $M_w$, and let us start at one of the leaf nodes. Figure~\ref{fig:node-deconstruction} shows the internal structure of each node for the expansion process. For the collapse process, we reverse the flow of time (swap the indices $i \leftrightarrow \left[i+1\right]$) and go from right to left in the circuit of Fig.~\ref{fig:decomposed-node}, taking the complex conjugate of $U_{\mathrm{ent}}$, $K_{m_r}$, and $K_{m_s}$. The collapse process tells us to use $\rho_{0,r} = \mathbb{I}/2 = \rho_{0,s}$ as inputs on the right of the diagram in Fig.~\ref{fig:decomposed-node}. Since we know $m_r$ and $m_s$ from $M_w$, we can efficiently compute $\rho_{1,q}$, because these operations only involve two qubits. Repeat this process for all the leaf nodes. Each node yields a new quantum state, so after completing the computation for the $2^{t-1}$ leaf nodes, the result is a collection of states $\Omega_1 \equiv \left\{\rho_{1,j}\right\}_{j=1}^{2^{t-1}}$. Use these as inputs to the next set of nodes at time step $i = 2$. Note that $2\lvert \Omega_i \rvert = \lvert \Omega_{i-1} \rvert$. Continue in this manner until one computes the final state, which is $\rho_t$ (in Fig.~\ref{fig:collapse-tree-diagram}, $t = 4$). One can then compute the components of $\vec{n}_t$ using $n_t^j = \Tr{\left[ \sigma_j \rho_t \right]}$ for $j \in \left\{x,y,z\right\}$.

\subsection{Circuit construction} \label{sec:circuit-construction}

Suppose we are given a time $t$ and a measurement strength $\theta$. To construct a circuit encoding an instance of the expansion process for the dynamical quantum tree model, we start by applying a Hadamard gate to the root qubit (initialized to $\ket{0}$), followed by a measurement in the computational basis (which we store as $m_0$). Such a procedure initializes the root qubit to $\ket{0}$ or $\ket{1}$ with equal probability. As we discussed in Section~\ref{sec:probe}, doing so yields the same probabilities for the measurement record $\mathcal{M}$ on line~\ref{line:sample-circuit} of Algorithm~\ref{alg:estimate_Z} as using a probe and root qubit beginning in the state $\ket{\Phi^+}$.

Next, we add the circuit components for each node in Fig.~\ref{fig:quantum-tree-diagram} by traversing the dynamical quantum tree in increasing time and from left to right at a given time. For each node, we apply the entangling circuit in Fig.~\ref{fig:entangling-circuit} with $\rho$ being the qubit from the previous time step (or the root qubit for $t = 1$). Note that for each entangling circuit, we sample new single-qubit unitaries $U_1$, $U_2$, $U_3$, and $U_4$ from the Haar measure. Every entangling circuit also requires a fresh ancillary qubit in state $\ket{0}$ which then becomes part of the dynamical quantum tree. Following the entangling circuit, we perform a weak measurement with measurement strength $\theta$ using the right circuit of Fig.~\ref{fig:weak-measurement-circuit} on each qubit that is output in the circuit of Fig.~\ref{fig:entangling-circuit}. To increase the weak measurement operation's fidelity, we use the circuit with an entangling gate native to the quantum computer we used for our experiment (right circuit of Fig.~\ref{fig:weak-measurement-circuit}). We store these measurement outcomes. Each weak measurement circuit requires an ancillary qubit in state $\ket{0}$, but these qubits are reusable because the projective measurement in the weak measurement circuit disentangles them from the rest of the qubits. For a general $t$ and some number $l$ of ancillary qubits we use to implement the weak measurements, we require $2^t + l$ qubits. There is one projective measurement (for the root qubit) and $2(2^{t}-1)$ weak measurements.

\subsection{Experimental details} \label{sec:experimental-details}

For our experiment, we used the 20-qubit Quantinuum H1-1 trapped-ion quantum computer~\cite{quantinuum2022}. Since we require $2^t+l \leq 20$, we chose $t = 4$ and $l = 4$, the largest $t$ we could implement on the H1-1. Note that taking $l > 1$ allows us to reduce the circuit depth by potentially parallelizing some of the weak measurements.  We submitted the circuits using the tket software library~\cite{tket2020}, and we optimized them using the flag \verb|tket_opt_level = 2|.

After executing a circuit for a given $t$, $\theta$, and $\mathcal{U}$, we obtain a measurement record $\mathcal{M} = (m_0, m_1, \ldots)$. Notice though how the circuit encoding an instance of the time $t$ expansion process for the dynamical quantum tree model has a recursive structure: the circuit contains sub-circuits encoding instances for the $t' < t$ expansion process for the dynamical quantum tree model. Therefore, one can truncate $\mathcal{U}$ and $\mathcal{M}$ to get results for $t' < t$. Concretely, we replace the results in lines~\ref{line:sample-unitaries} and~\ref{line:sample-circuit} of Algorithm~\ref{alg:estimate_Z} by values derived from $\mathcal{U}$ and $\mathcal{M}$ for the circuits encoding instances at time $t$. In our case, we only execute circuits for $t = 4$. Let $\mathcal{U}'$ be the list of unitaries from $\mathcal{U}$ describing an instance for the $t' < t =4$ expansion process for the dynamical quantum tree model, and let $\mathcal{M}'$ be the measurement record obtained by truncating $\mathcal{M}$ to only its first $1 + 2(2^{t'}-1)$ elements. Then, one can substitute $\mathcal{U}'$ and $\mathcal{M}'$ in lines~\ref{line:sample-unitaries} and~\ref{line:sample-circuit} of Algorithm~\ref{alg:estimate_Z} to create an estimate for $\hat{Z}_{t',N_c}(\theta)$. We do this to produce the $t' \in \left\{1,2,3\right\}$ data in Fig.~\ref{fig:experimental-results}, though we note that doing so introduces dependencies between the datasets for different values of $t$.

To estimate the resources we consumed for our experiment, we use Quantinuum's concept of a H-System Quantum Credit (HQC). One can extract the HQC cost directly from any circuit, since the cost depends only on the number of single-qubit gates, the number of two-qubit gates, the number of operations related to state preparation and measurement, and the number of shots for the circuit. Given $t$, $\theta$ and $\mathcal{U}$, we constructed a circuit encoding an instance of the expansion process for the dynamical quantum tree model as described in Section~\ref{sec:circuit-construction}. We then packaged several instances together by concatenating the circuits (separated by qubit resets) to create a deeper circuit for execution on the Quantinuum H1-1. Logically, there is no difference between running the circuits individually or as one concatenated circuit. However, the latter requires fewer HQCs to execute, allowing us to run more circuits for our fixed budget. In our experiment, we concatenated $5$-$6$ circuits at a time. We also made the simplifying assumption in our analysis that the measurement outcomes between instances of a concatenated circuit are independent of each other. We find our $994$ circuits consumed a total of approximately $13 099$ HQCs, of which we had an allocation of $11000$ HQCs from Oak Ridge Leadership Computing Facility. Using that allocation, we ran $834$ circuits during the period of October 08-24, 2024, and Quantinuum reported to us that they took approximately $402$ minutes to run (or approximately $0.482$ minutes per circuit). Quantinuum ran the remaining $160$ circuits for us separately, but we do not have timing data for them. However, since their structure is similar to the other circuits, we use $0.482$ minutes per circuit as a proxy for the running time. We therefore find that the approximate total running time for all $994$ circuits is $480$ minutes, or $8$ hours.

For six values of $\theta$ and $t = 4$, we collected data for $N_c = 834$ instances of the expansion process for the dynamical quantum tree model, with $N_s = 8$ shots per instance. For $t' < t =4$, we collected data as we described above. We summarize all the data in Section~\ref{sec:results} by considering, for a given $t$ and $\theta$, the estimate $\hat{Z}_{t,N_c}(\theta) \pm z E_{t,N_c}(\theta)$, with $z = 1.96$ representing the $z$-score for a $95\%$ confidence interval and $E_{t,N_c}(\theta)$ being the standard error of the mean for $\hat{Z}_{t,N_c}(\theta)$, as defined on line~\ref{line:error} in Algorithm~\ref{alg:estimate_Z}. Note that we did not perform any error-mitigation of the results; we only analyzed them to estimate the order parameter. We include the measurement outcomes and results, as well as code to generate and submit the $994$ total circuits, in a public Zenodo repository~\cite{experimental-data}.

\subsection{Calculating the order parameter for the expansion process using a modified collapse process}
\label{appendix:reverse_map}

In the main text, we focused on the MIPT in the expansion process of the dynamical quantum tree model. Despite being able to use a modified version of the analytical approach in ref.~\cite{Feng_charge_2024} for the MIPT, we also want to numerically simulate the dynamics for finite $t$ so we can compare experiment to theory (the dashed curves in Fig.~\ref{fig:experimental-results}). Unfortunately, classically simulating the dynamics of a system with size $\mathcal{N} = 2^t$ via brute-force can be challenging as $t$ increases (see ref.~\cite{Feng_charge_2024} for more details). Although our experimental study of $t \leq 4$ makes brute-force simulation possible, we wish to obtain the order parameter using a numerical method that can easily scale to very large system sizes. To circumvent this challenge in brute-force simulation, we begin with the mapping in Section~\ref{sec:reverse-construction-bloch-vector}, which allows us to calculate the probe qubit's state in the expansion process for a given $\mathcal{U}$ and $M_w$ by computing the final state in the collapse process of the dynamical quantum tree model. Furthermore, the computation of the final state in the collapse process is efficient. Unfortunately, to classically simulate the Kraus operators (and therefore $M_w$) necessary to compute the final state in the collapse process, we require knowledge of the entire dynamical quantum tree at earlier times. This is inefficient in $\mathcal{N}$. We trace the inefficiency to the requirement that we post-select one of the output states in each node of Fig.~\ref{fig:decomposed-node} to be $\ket{0}\bra{0}$~\cite{Feng_charge_2024}. We note again that, as in Section~\ref{sec:reverse-construction-bloch-vector}, each node in the collapse process corresponds to the `complex conjugate' circuit of Fig.~\ref{fig:decomposed-node}. That is, reflect the circuit about the horizontal ($y$-axis), reverse the flow of time (swap the indices $i \leftrightarrow [i+1]$), and take the complex conjugate of all circuit components. Such a procedure makes $\ket{0}\bra{0}$ one of the output states. For the rest of the paper, whenever we refer to Fig.~\ref{fig:decomposed-node}, we have in mind the complex conjugated versions of these operators, as we just described.

Suppose we make a change to the node in Fig.~\ref{fig:node-deconstruction}: instead of post-selecting one of the output states to be $\ket{0}\bra{0}$, we simply measure that qubit in the computational basis. Let $M_p$ be the resulting projective measurement outcomes, and we define $M_p^0$ to our original setting where every projective measurement yields outcome $0$. At first glance, such a change does not seem to yield much. However, we only care about computing the expectation value of an arbitrary function of the probe qubit's state $f(\rho_t(\theta, \mathcal{U}, M_p^0, M_w))$, where we now make the dependence of the probe qubit's state on $M_p^0$ explicit. Suppose we replace the postselection $M_p = M_p^0$ by sampling $M_p$ according to Born's rule and averaging over $M_p$ as well as $\mathcal{U}$ and $M_w$. We find that averaging the function over $M_p$ (according to Born's rule), $\mathcal{U}$, and $M_w$ gives the same result as $\mathbb{E}_{\mathcal{U}, M_w}\left[f(\rho_t(\theta, \mathcal{U}, M_p^0, M_w))\right]$, while also admitting easy classical simulation for large system sizes~\cite{nahum_measurement_2021,feng_measurement_2023}, which we describe in Section~\ref{appendix:pool_method}.

We begin by noting an invariance property of the probe qubit's state. Suppose $t$, $\theta$, and $M_w$ are fixed. Then, for any pair $(M_p, \mathcal{U})$, we define an equivalent new pair $(M_p^0, \mathcal{U}_{M_p})$, where we obtain $\mathcal{U}_{M_p}$ in the following way: for any $m \in M_p$, we multiply the associated unitary (corresponding to $U_2^\dagger$ in the complex conjugated circuit of Fig.~\ref{fig:entangling-circuit}) in $\mathcal{U}$ by $X^m$ (using the Pauli $X$ gate). We interpret such a transformation as `absorbing' the projective measurements into $\mathcal{U}$. Since the transformation is only a redefinition of the unitary operators in $\mathcal{U}$, we find:
\begin{equation} \label{eq:rho-invariance}
    \rho_t(\theta,\mathcal{U}, M_p, M_w) = \rho_t(\theta,\mathcal{U}_{M_p}, M_p^0, M_w).
\end{equation}
Using this result, we find that the expectation value of the function $f$ does not change if we condition on a particular value of $M_p^0$:
\begin{align} \label{eq:f-expctation-invariance}
    \mathbb{E}_{\mathcal{U}, M_p, M_w} &\left[f(\rho_t(\theta,\mathcal{U}, M_p, M_w))\right] \notag \\
    &= \mathbb{E}_{\mathcal{U}, M_p, M_w} \left[f(\rho_t(\theta,\mathcal{U}_{M_p}, M_p^0, M_w))\right] \notag \\
    &= \mathbb{E}_{\mathcal{U}, M_p, M_w} \left[f(\rho_t(\theta,\mathcal{U}, M_p^0, M_w))\right] \notag \\
    &= \mathbb{E}_{\mathcal{U}, M_w} \left[f(\rho_t(\theta,\mathcal{U}, M_p^0, M_w))\right].
\end{align}
The first equality uses equation~\eqref{eq:rho-invariance}. We get the second equality by noticing that averages over $\mathcal{U}$ (and therefore the Haar measure for single-qubit unitaries) are invariant over single-qubit unitary transformations. The final equality simply recognizes that $M_p$ does not appear in the expectation value and is therefore irrelevant.

We conclude that we can obtain $Z_t(\theta)$ and $Z_t^{\mathrm{typ}}(\theta)$ for the expansion process of the dynamical quantum tree model by considering the corresponding  collapse process of the dynamical quantum tree model where we sample $M_p$ and $M_w$ through Born's rule. Refs.~\cite{nahum_measurement_2021,feng_measurement_2023,Monthus_Anderson_2009,Garcia-Mata_Scaling_2017,Shi_Classical_2006,Murg_Simulating_2010,Silvi_Homogeneous_2010,Tagliacozzo_Simulation_2009,Li_Efficient_2012} showed how to solve and simulate the collapse process of the dynamical quantum. We can translate these results to our setup, allowing us to study the phase transition in Section~\ref{sec:tree_model} analytically and numerically. In Section~\ref{appendix:theory} we summarize the theory of the collapse process of the dynamical quantum tree model where we sample $M_p$ and $M_w$ through Born's rule. In Section~\ref{appendix:pool_method}, we discuss our method for numerically obtaining the simulation results in Fig.~\ref{fig:experimental-results}. 

\subsection{The phase transition for the collapse process of the dynamical quantum tree model}
\label{appendix:theory}
Previous work~\cite{nahum_measurement_2021,feng_measurement_2023} established the detailed theory of the phase transition for the collapse process of the dynamical quantum tree model. Here, we review the main parts and specialize to our particular ensemble of unitary gates (since previous work used two-qubit Haar-random unitary gates instead of our entangling circuit in Fig.~\ref{fig:entangling-circuit}). For more details on the methods and derivations we use in this section, see Section~IV of ref.~\cite{nahum_measurement_2021} and Section~IV of ref.~\cite{feng_measurement_2023}.

Consider a collapse process for the dynamical quantum tree up to time $t$, as in Fig.~\ref{fig:collapse-tree-diagram}. The initial state is the maximally mixed state on $\mathcal{N} = 2^t$ qubits, which we get by considering the input states to the leaves of the tree in Fig.~\ref{fig:collapse-tree-diagram}: $\rho_0^{\otimes \mathcal{N}} = \left(\mathbb{I}/2\right)^{\otimes 2^t} = \mathbb{I}_{2^t}/2^{2^t}$. As we describe in Section~\ref{sec:reverse-construction-bloch-vector} and show in Fig.~\ref{fig:node-deconstruction}, each node of the collapse process in the dynamical quantum tree model involves first the application of individual weak measurements on the input qubits to a given node, followed by an entangling circuit and a projective measurement. In Section~\ref{sec:reverse-construction-bloch-vector} we post-select the outcome of the projective measurement to be $0$, but in Section~\ref{appendix:reverse_map} we showed that expectation values for arbitrary functions $f(\theta, \mathcal{U}, M_w)$ are unchanged even if we do not post-select the projective measurements and simply average over them as well. Let $M$ be the list of weak and projective measurement outcomes after executing the circuit encoding an instance of the dynamical quantum tree model, excluding the measurement outcome for the probe qubit. Since a projective measurement disentangles a qubit from the others, we will ignore it for the rest of the evolution and only focus on the qubits we did not projectively measure. In this way, each time step reduces the number of qubits we keep track of by half; after $t$ steps, a single qubit of interest with state $\rho_t(\theta, \mathcal{U}, M)$ remains. We can write the state as:
\begin{equation}
    \rho_t(\theta,\mathcal{U},M) = \nu
    \begin{pmatrix}
    1-Z_t(\theta,\mathcal{U},M) &0\\
    0& Z_t(\theta,\mathcal{U},M)
    \end{pmatrix}\nu^{\dagger},
\end{equation}
with $\nu$ as the basis, which generally depends on $t$, $\theta$, $\mathcal{U}$ and $M$ but we ignore them for brevity. Since we are only concerned with quantities that are averages over $\mathcal{U}$ and $M$, we can use the Haar-invariance of the unitaries in $\mathcal{U}$ to absorb the basis of $\rho_t(\theta, \mathcal{U}, M)$ into the entangling circuit associated to the top-most node in Fig.~\ref{fig:collapse-tree-diagram}. As in Section~\ref{sec:tree_model}, we introduce the smaller eigenvalue $Z_t(\theta, \mathcal{U}, M)$ of $\rho_t(\theta, \mathcal{U}, M)$ to fully characterize the dynamics. Furthermore, previous work~\cite{nahum_measurement_2021,feng_measurement_2023} showed that the typical value of $Z_t(\theta, \mathcal{U}, M)$ acts as an order parameter for the model, defined by:
\begin{equation}
\ln Z^{\mathrm{typ}}_t(\theta)\equiv \mathbb{E}_{\mathcal{U},M} [\ln Z_t(\theta,\mathcal{U},M)].
\end{equation}
As in Section~\ref{sec:tree_model}, the model has two phases. The mixed phase occurs when $\theta \in \left[\pi/2, \theta_c\right)$, where $\ln Z^{\mathrm{typ}}_t(\theta)$ remains finite even in the limit $t\to \infty$. The pure phase occurs when $\theta \in \left(\theta_c, \pi\right]$, where $Z^{\mathrm{typ}}_t(\theta)$ decays to $0$ as $t \to \infty$.

Due to the recursive structure of the dynamical quantum tree model, we can write down a recursion relation for $Z_t(\theta,\mathcal{U},M)$ in terms of the previous time step $t-1$. Graphically, one can associate a value $Z_i(\theta, \mathcal{U}, M)$ to any node at time $i\leq t$ in Fig.~\ref{fig:collapse-tree-diagram} by viewing that node as the root of its own quantum tree. We call these \emph{sub-trees}, and one can write the recursion relation for $Z_t(\theta,\mathcal{U},M)$ at time $t$ in terms of $Z'_{t-1}(\theta,\mathcal{U},M)$ and $Z''_{t-1}(\theta,\mathcal{U},M)$, corresponding to the two sub-trees connected to the root node for the collapse process of the dynamical quantum tree model at time $t$. In general, the recursion relation is nonlinear and complicated to solve. However, previous work proved~\cite{nahum_measurement_2021,feng_measurement_2023} that only the linear terms of the recursion relation play a role in determining the critical point $\theta_c$ and the scaling behavior of $Z_t(\theta, \mathcal{U}, M)$ near $\theta_c$. For a specific realization of the collapse process for the dynamical quantum tree model up to time $t$, we can write the recursion relation as:
\begin{equation} \label{eq:recursion-relation}
    Z_t = A_1Z'_{t-1}+A_2Z''_{t-1}+\mathcal{O}(Z_{t-1}^{\prime 2},Z'_{t-1}Z''_{t-1},Z^{\prime\prime 2}_{t-1}),
\end{equation}
where we suppress all arguments for these terms to simplify the notation in this equation. We note that $A_1 = A_1(\theta, \mathbf{U}, \mathbf{M})$ and $A_2 = A_2(\theta, \mathbf{U}, \mathbf{M})$, where $\mathbf{U}$ is the choice of unitary gates at the root node of the collapse process for the dynamical quantum tree at time $t$ and $\mathbf{M}$ are the measurement outcomes for the same node. One can also write a similar relation to equation~\eqref{eq:recursion-relation} for any node in the dynamical quantum tree in terms of its sub-trees. In the collapse process, every node yields the same statistics for $A_1(\theta, \mathbf{U}, \mathbf{M})$ and $A_2(\theta, \mathbf{U}, \mathbf{M})$. To see why this is true, we first note that every node has the same value for $\theta$ and $\Pr[\mathbf{U}]$. To determine the statistics of $A_1$ and $A_2$, we also need to get the statistics of $\mathbf{M}$ in the node. In principle, the statistics of $\mathbf{M}$ depend on the inputs $Z'_{t-1}(\theta,\mathcal{U},M)$ and $Z''_{t-1}(\theta,\mathcal{U},M)$. This should lead to different statistics of $A_1$ and $A_2$ in different nodes. However, ref.~\cite{feng_measurement_2023} found that one can extract information about the phase transition despite ignoring the effects of the inputs. This guarantees that the statistics of the coefficients $A_1(\theta, \mathbf{U}, \mathbf{M})$ and $A_2(\theta, \mathbf{U}, \mathbf{M})$ are the same for every node.

As shown in previous work~\cite{Derrida_Polymers_1988,nahum_measurement_2021,feng_measurement_2023}, the statistics of coefficients $A_1(\theta,\mathbf{U},\mathbf{M})$ and $A_2(\theta,\mathbf{U},\mathbf{M})$ fully determine the dynamics of $Z_t(\theta,\mathcal{U},M)$ near the critical point. To see this, we first introduce a generating function $G_t(x,\theta) = \mathbb{E}_{\mathcal{U},M} [\exp(-e^{-x}Z_t(\theta,\mathcal{U},M))]$ to help describe the evolution of statistics of $Z_t(\theta,\mathcal{U},M)$. If we interpret $x$ as position and $t$ as the time, then $G_t(x,\theta)$ describes a moving wave from right to left, with the wavefront occurring at $\ln Z^{\mathrm{typ}}_t(\theta)$. Using the definition of $G_t(x, \theta)$, we can write down an analogous recursion relation to equation~\eqref{eq:recursion-relation} for $G_t(x,\theta)$, which is the discrete version of the Fisher-KPP equation. In the pure phase, we can rewrite $G_t(x,\theta)$ in the form $G_t(x-v_{\theta}t)$ where $v_{\theta}$ is the velocity of the moving wavefront $\ln Z_t^{\mathrm{typ}}(\theta)$ as $t\to\infty$. In the limit $(x-v_{\theta}t)\to\infty$, we can approximate the generating function as $G_t(x-v_{\theta}t) \approx 1-\exp\left\{-\lambda(x-v_{\theta}t)\right\}$ with $\lambda$ being a parameter determined by the stability of $G_t(x-v_{\theta}t)$~\cite{Derrida_Polymers_1988}. Substituting the approximation into the recursion relation of equation~\eqref{eq:recursion-relation}, we can write a recursion relation for $G_t(x,\theta)$:
\begin{equation}
    G_{t}(x,\theta)=\mathbb{E}_{\mathbf{U},\mathbf{M}}\left[G_{t-1}(x-\ln A_1,\theta)G_{t-1}(x-\ln A_2,\theta))\right],
\end{equation}
where we suppress the arguments of $A_1$ and $A_2$ in this equation for simplicity. We can then derive the velocity:
\begin{equation} \label{eq:velocity}
    v_{\theta} =\frac{1}{\lambda}\ln \left(\mathbb{E}_{\mathbf{U},\mathbf{M}}[ A_1^{\lambda}(\theta,\mathbf{U},\mathbf{M})+ A_2^{\lambda}(\theta,\mathbf{U},\mathbf{M})]\right).
\end{equation}

In the pure phase ($\theta > \theta_c$), we find $\ln Z_t^{\mathrm{typ}}(\theta) \sim v_{\theta}t$ as $t\to\infty$. In the mixed phase, the quantity $\ln Z^{\mathrm{typ}}_t(\theta)$ saturates when $t\to\infty$ . This tells us that the velocity vanishes at the critical point. Thus, we can determine the critical point by finding the value of $\theta$ for which $v_{\theta}=0$. Furthermore, we can determine the parameter $\lambda$ by solving $\partial_{\lambda}v_{\theta}=0$, a stability condition for $G_t(x,\theta)$. For our setup with outcomes of the weak and projective measurements obeying Born's rule, $\lambda =1$ at the critical point~\cite{feng_measurement_2023}. Using this result, we obtain the critical point $\theta_c$ by numerically solving the equation:
\begin{equation}
    \mathbb{E}_{\mathbf{U},\mathbf{M}}\left[A_1(\theta,\mathbf{U},\mathbf{M})+A_2(\theta,\mathbf{U},\mathbf{M})\right]=1.
\end{equation}
We find $\theta_c = 2.2142(2)$. Besides the critical point, we can further derive the universality class of the phase transition, using the same method in ref.~\cite{nahum_measurement_2021,feng_measurement_2023}. We get:
\begin{align}
     Z^{\mathrm{typ}}_{t\to\infty}(\theta)& \sim \exp\left\{-\frac{C}{\sqrt{\theta_c-\theta}}\right\} & (\theta & \lesssim \theta_c ), \notag\\
     \ln Z^{\mathrm{typ}}_t(\theta)& \sim -t^{1/3} & (\theta & =\theta_c),
     \label{eq:scale}
\end{align}
where $C$ is a real, positive constant that depends on the statistics of $A_1(\theta,\mathbf{U},\mathbf{M})$ and $A_2(\theta,\mathbf{U},\mathbf{M})$. Note that the exact value of $C$ is unknown (see ref.~\cite{feng_measurement_2023}). Furthermore, refs.~\cite{nahum_measurement_2021,feng_measurement_2023} showed that when $\theta \lesssim \theta_c$, the quantities $Z_{t\to\infty}(\theta)=\mathbb{E}_{\mathcal{U},M}\left[Z_{t\to\infty}(\theta,\mathcal{U},M)\right]$ and $Z_{t\to\infty}^{\mathrm{typ}}(\theta)$ satisfy:
\begin{equation}
    Z_{t\to\infty}(\theta)\sim \left(Z^{\mathrm{typ}}_{t\to\infty}(\theta)\right)^{\lambda}.
\end{equation}
Since $\lambda = 1$ in our case, we have $Z_{t\to\infty}(\theta)\sim Z_{t\to\infty}^{\mathrm{typ}}(\theta)$. This allows us to replace $Z_t^{\mathrm{typ}}(\theta)$ by $Z_{t}(\theta)$ to characterize the phase transition.

\subsection{Pool method}
\label{appendix:pool_method}
\begin{figure*}
    \centering
    \includegraphics[width=0.9\textwidth]{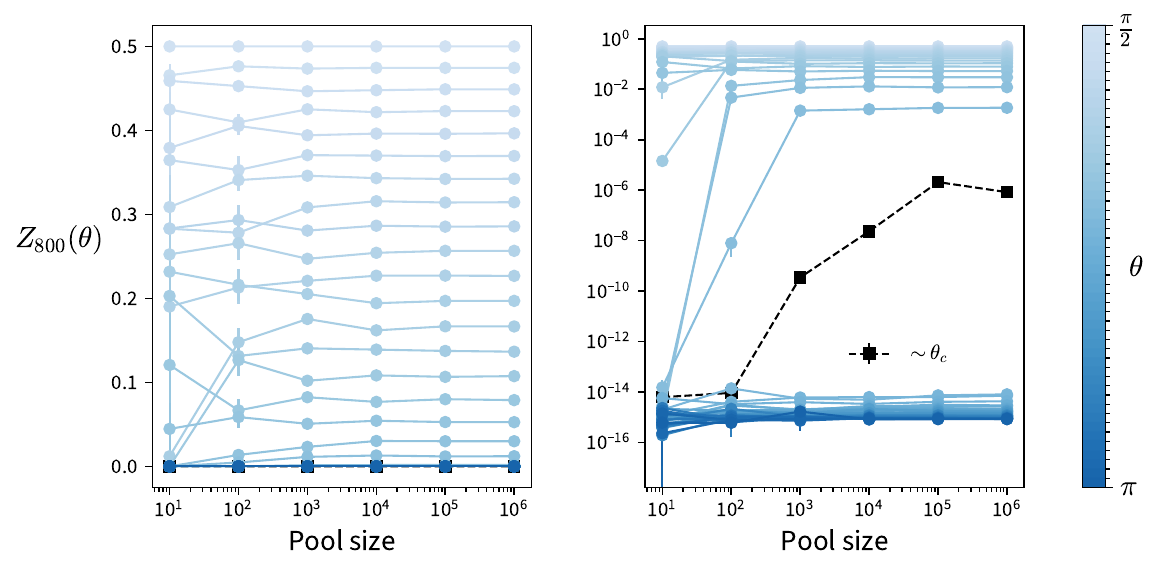}
    \caption{\justifying Convergence of $Z_t(\theta)$ as a function of pool size for the pool method, with $t = 800$ and various measurement strengths (shades of blue for the data). The two panels display the same data; the left panel uses a linear vertical scale and the right panel uses a logarithmic vertical. The data are the markers, with the connecting lines simply a guide for the eye. We highlight the data closest to the critical point with black squares and dashed lines. Note that both panels use a logarithmic horizontal scale. In the left panel, we observe that $Z_{800}(\theta)$ stabilizes as a function of pool size for all angles. In the right panel, we see that deviations near the critical point $\theta_c$ span several orders of magnitude but still remain close to zero ($\sim 10^{-6}$). We also plot the $\sim \theta_c$ data in the left panel, though the values are all nearly zero, so they at the bottom of the panel. Error bars for each data point show a $95\%$ confidence interval ($1.96 \cdot \mathrm{SE}$, where $\mathrm{SE}$ is the standard error of the mean for a given pool size). Note that many error bars are smaller than the marker width.}
    \label{fig:convergence-results}
\end{figure*}

For the problem of the dynamical quantum tree model, one important numerical method is called the \textit{pool method}~\cite{Miller_Weak-disorder_1994,Monthus_Anderson_2009,Garcia-Mata_Scaling_2017,Shi_Classical_2006,Murg_Simulating_2010,Silvi_Homogeneous_2010,Tagliacozzo_Simulation_2009,Li_Efficient_2012}. The main idea is that, as in Section~\ref{appendix:theory}, we can view the dynamical quantum tree in the collapse process as connecting two sub-trees to a single node. For instance, in Fig.~\ref{fig:collapse-tree-diagram}, we can think of the root node for the $t = 4$ tree as connecting two sub-trees whose roots are located at $t = 3$ (there are two of them). Using the probability distribution of $\rho_{t-1}(\theta,\mathcal{U},M)$, we can obtain the probability distribution of $\rho_t(\theta,\mathcal{U},M)$. In what follows, we will use $M$ to generically denote the measurement outcomes for the collapse tree up to some time; so technically the $M$ associated to $\rho_{i-1}(\theta,\mathcal{U},M)$ and $\rho_i(\theta,\mathcal{U},M)$ will differ. To do this numerically, we establish a pool to store $Z_t(\theta,\mathcal{U},M)$ values for every $t$. This pool acts as an approximate distribution of $Z_t(\theta,\mathcal{U},M)$ for the collapse process of the dynamical quantum tree model at time $t$. Suppose now we have the pool for time $(t-1)$. We can compute the pool for time $t$ as follows. First, we choose two $Z_{t-1}(\theta,\mathcal{U},M)$ values from the pool uniformly at random and construct the corresponding density matrices as in equation~\eqref{eq:rho_k}. As we discussed in Section~\ref{appendix:reverse_map}, expectation values for functions of the density matrix are invariant under single-qubit unitary rotations of any $U \in \mathcal{U}$. This allows us to give each $\rho_{t-1}(\theta,\mathcal{U},M)$ a Haar-random basis $\nu$ as in equation~\eqref{eq:rho_k} without loss of generality. Second, for each $\rho_{t-1}(\theta,\mathcal{U},M)$ we sample a weak measurement operator $K_m^{\dagger}=K_m$ from equation~\eqref{eq:kraus-operator} with outcome $m \in \left\{0,1\right\}$ according to the probability $\Pr\left[m\right] = \Tr\left[K_{m}\rho_{t-1}(\theta,\mathcal{U},M)K^{\dagger}_{m}\right] =\Tr\left[K^{\dagger}_{m}\rho_{t-1}(\theta,\mathcal{U},M)K_{m}\right]$. We then apply the operators to the density matrices and record the measurement outcomes. Third, we sample four single-qubit Haar-random unitaries, use them to construct the entangling circuit as in Fig.~\ref{fig:entangling-circuit} and then use the two density matrices as input. Finally, we projectively measure one of the qubits at the output, record the measurement outcome, and discard that qubit. Using the decomposition of equation~\eqref{eq:rho_k}, we calculate the $Z_t(\theta,\mathcal{U},M)$ value for the state of the qubit we did not measure, and then we add that value to the pool for time $t$. We repeat the above steps for some fixed pool size, with the goal of taking sufficiently many values such that the pool approximates the distribution of $Z_t(\theta,\mathcal{U},M)$. To begin the pool method, we start with a pool determined by the initial state of the collapse process for the dynamical quantum tree model and recursively construct the rest of the pools for each time step. For each $t$, we calculate the order parameter by averaging the $Z_t$ values in the final pool corresponding to the root node of the dynamical quantum tree. We can also calculate the standard error of the mean $\mathrm{SE}$ using the values in the pool. Previous work~\cite{nahum_measurement_2021} showed that the numerical results converge quickly with the size of the pool, roughly independent of $t$. We observe the same here for $t \leq 800$. In Fig.~\ref{fig:convergence-results}, we show the convergence of $Z_t(\theta)$ using the pool method for $t = 800$ and all the measurement strengths we consider in Fig.~\ref{fig:experimental-results}. We find that $Z_{800}(\theta)$ (for any $\theta$ we considered) stabilizes for pool sizes of around $10^5-10^6$. Since we observe convergence for $t = 800$, we expect the same to be true for smaller $t$. Because we established a mapping between the expansion process of the dynamical quantum tree model (our experiment) and the collapse process of the dynamical quantum tree model where we replace the postselection with projective measurements, the pool method allows us to easily simulate the results of our experiment to compare against.

\begin{figure}
    \centering
    \includegraphics[width=1\columnwidth]{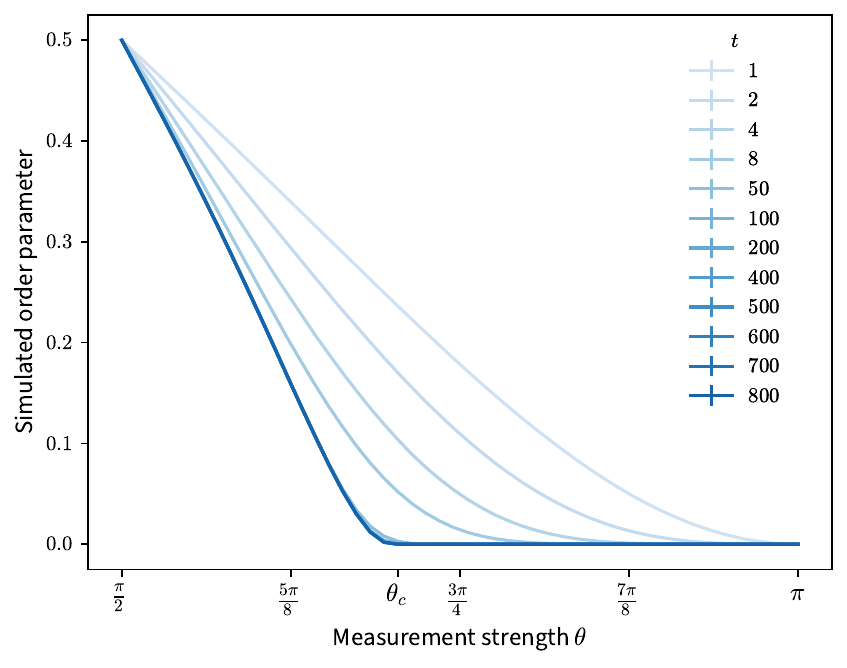}
    \caption{\justifying Convergence of the curves $Z_t(\theta)$ to $Z_{t\to\infty}(\theta)$ for different values of $t$ using the pool method with pool size $10^6$. Increasing values of $t$ correspond to darker blue. Given $t$, data points are $(\theta, Z_t(\theta))$ for $50$ different measurement strengths $\theta$. For clarity, we omit the markers for the data points and only show the connecting lines. We do include error bars for each data point corresponding to a $95\%$ confidence interval for $Z_t(\theta)$. We use error bars of size $1.96 \cdot \mathrm{SE}$, where $\mathrm{SE}$ is the standard error of the mean of $Z_t(\theta)$ using the pool method, though these error bars are negligible compared to the line width.}
    \label{fig:simulated-order-parameter}
\end{figure}

We also claim that the curve $Z_{800}(\theta)$ we obtained from the pool method is a good proxy for $Z_{t\to\infty}(\theta)$ in Fig.~\ref{fig:experimental-results}. As we show in Fig.~\ref{fig:simulated-order-parameter}, the curves for $Z_t(\theta)$ using the pool tend towards a characteristic form as $t$ increases. In particular, we see that the $t = 200$ curve is already comparable to the $t = 800$ curve. While we do not have a proof that the pool method is efficient in $\mathcal{N} = 2^t$ since we do not know how the pool size required for convergence scales with $t$, we can obtain simulation results for up to $t = 800$. Such an experiment would require a galactic $\mathcal{N} = 2^{800}$ qubits, a practical impossibility. Therefore, we conclude that the pool method allows us to simulate the value of $Z_t(\theta)$ and compare it against any practical experiment. In Fig.~\ref{fig:experimental-results}, the numerical simulation of the dashed and dot-dashed curve uses the pool method with pool size $10^6$. We find the maximum standard error of the mean for all $50$ measurement strengths considered here and $t \in \left\{1,2,\ldots, 800\right\}$ is less than $1.04 \cdot 10^{-4}$, so the corresponding $95\%$ confidence interval with $z = 1.96$ is less than $2.1 \cdot 10^{-4}$, which is negligible compared to the line width.

\subsection{Equivalence of weak measurement circuits} \label{sec:weak-measurement-circuits}

To show the equivalence of both circuits in Fig.~\ref{fig:weak-measurement-circuit}, we demonstrate that the output state before measurement is identical up to a global phase. Instead of working with density matrices as in Fig.~\ref{fig:weak-measurement-circuit}, we will instead define the pure state $\ket{\psi} = a \ket{u}_A\ket{0}_B + b \ket{v}_A \ket{1}_B$, with register $A$ encoding some multi-qubit states $\ket{u}_A$ and $\ket{v}_B$, normalized so $\lvert \braket{\psi|\psi} \rvert^2 = 1$. Note that the $B$ register represents a single qubit, and that is the qubit which we input into the circuits of Fig.~\ref{fig:weak-measurement-circuit}. Beginning with the left circuit in Fig.~\ref{fig:weak-measurement-circuit} (with identity acting on the qubits in $A$), we find:
\begin{align} \label{eq:left-circuit-state}
    \ket{\psi_f}&\equiv \mathrm{CX} \cdot R_Y(\theta) \ket{\psi} \ket{0} \notag \\
    & = \mathrm{CX} \ket{\psi} \left( \cos{\left(\theta/2 \right)} \ket{0} + \sin{\left(\theta/2 \right)} \ket{1} \right) \notag \\
    & = a \cos{\left(\theta/2 \right)} \ket{u}_A \ket{0}_B \ket{0} + a \sin{\left(\theta/2 \right)} \ket{u}_A \ket{0}_B \ket{1} \notag \\
    &+ b \sin{\left(\theta/2 \right)} \ket{v}_A \ket{1}_B \ket{0} + 
    b \cos{\left(\theta/2 \right)} \ket{v}_A \ket{1}_B \ket{1}.
\end{align}
Note that the $R_Y(\theta)$ gate only acts on the ancillary qubit, while the $\mathrm{CX}$ gate generates entanglement between the qubit in register $B$ and the ancillary qubit. Now consider the right circuit of Fig.~\ref{fig:weak-measurement-circuit}. We find the state evolves as:
\begin{align} \label{eq:right-circuit-state}
    \ket{\psi'_f} &\equiv R_X\left(\frac{\pi}{2}\right) \cdot ZZ(\phi) \ket{\psi} \ket{+} \notag \\
    &= \frac{R_X(\frac{\pi}{2})}{\sqrt{2}} \left[ a \mathrm{e}^{-i\frac{\phi}{2}} \ket{u}_A \ket{0}_B \ket{0} + a \mathrm{e}^{i \frac{\phi}{2}} \ket{u}_A\ket{0}_B \ket{1} \right. \notag \\
    &\left. \phantom{\frac{R_X(\frac{\pi}{2})}{\sqrt{2}} \left[\right.} + b \mathrm{e}^{i \frac{\phi}{2}} \ket{v}_A \ket{1}_B \ket{0} + 
    b \mathrm{e}^{-i \frac{\phi}{2}} \ket{v}_A \ket{1}_B \ket{1}\right] \notag \\
    &= \frac{\left( a \mathrm{e}^{-i \frac{\phi}{2}} \ket{u}_A \ket{0}_B + b \mathrm{e}^{i \frac{\phi}{2}} \ket{v}_A \ket{1}_B\right) R_X\left(\frac{\pi}{2}\right) \ket{0}}{\sqrt{2}} \notag \\
    &+ \frac{\left( a \mathrm{e}^{i \frac{\phi}{2}} \ket{u}_A \ket{0}_B + b \mathrm{e}^{-i \frac{\phi}{2}} \ket{v}_A \ket{1}_B\right) R_X\left(\frac{\pi}{2}\right) \ket{1}}{\sqrt{2}} \notag \\
    &= \left( \frac{1-i}{\sqrt{2}}\right)\ket{\psi_{f}},
\end{align}
where the last equality uses the relation $\phi = \pi/2-\theta$ to simplify the complex exponential and trigonometric functions. Therefore, both circuits achieve the same unitary transformation (up to a global phase).

\section*{Data and code availability}
Data and code for this experiment is available at a public Zenodo repository~\cite{experimental-data}. We include the measurement outcomes and circuits we used for the experiment, code to create the circuits and analyze the data, as well as code for classically simulating the order parameter.

\acknowledgments{
We thank Michael Foss-Feig for assistance with running our circuits on the Quantinuum quantum computers and for suggesting the weak measurement circuit construction in Fig.~\ref{fig:weak-measurement-circuit}. We thank Zihan Cheng and Matteo Ippoliti for useful discussions. This work was supported by the Ministère de l’Économie et de l’Innovation du Québec via its contributions to its Research Chair in Quantum Computing and to the AlgoLab at Institut quantique, Université de Sherbrooke. SK and JC were supported by a Natural Sciences and Engineering Research Council of Canada Discovery grant. This research used resources of the Oak Ridge Leadership Computing Facility, which is a DOE Office of Science User Facility supported under Contract DE-AC05-00OR22725.}

\section*{Author contributions}
X.F. and J.C. contributed equally. X.F. and J.C. developed the code for the experiment and analyzed the results from numerical simulations and experiment. B.S. and S.K. supervised the project. All authors wrote the manuscript.

\bibliography{references}
\end{document}